\documentclass[]{aa} 
\usepackage[varg]{txfonts}
\usepackage{xcolor}
\usepackage{inputenc}
\usepackage[]{graphicx}
\begin{document}

\title{On the multiscale behaviour of stellar activity and rotation of the planet host Kepler-30}

\author{D. B. de Freitas\inst{1,2}
\thanks{Email: danielbrito@fisica.ufc.br}
  \and A. F. Lanza\inst{1}
  \and F. O. da Silva Gomes\inst{2}
  \and M. L. Das Chagas\inst{3}}

\institute{INAF-Osservatorio Astrofisico di Catania, Via S. Sofia 78, I-95123, Catania, Italy
  \and Departamento de F\'{\i}sica, Universidade Federal do Cear\'a, Caixa Postal 6030, Campus do Pici, 60455-900 Fortaleza, Cear\'a, Brazil
  \and Faculdade de F\'{\i}sica $\hat{\rm a}$ Instituto de Ci\^encias Exatas, Universidade Federal do Sul e Sudeste do Par\'a, Marab\'a, PA 68505-080, Brazil}

\date{}

\abstract{The Kepler-30 system consists of a  G dwarf star with a rotation period of $\sim$16 days and three planets  orbiting almost coplanar with periods ranging from 29 to 143 days. Kepler-30 is a unique target to study stellar activity and rotation in a young solar-like star accompanied by a compact planetary system.}{We use about 4 years of high-precision photometry collected by the Kepler mission to investigate the fluctuations caused by photospheric convection, stellar rotation, and starspot evolution as a function of the timescale. Our main goal is to apply methods for the analysis of timeseries to find the timescales of the phenomena that affect the light variations. We correlate those timescales with periodicities in the star as well as in the planetary system.}{We model the flux rotational modulation induced by active regions using spot modelling and apply the Multifractal Detrending Moving Average algorithm (MFDMA) in standard and multiscale versions for analysing the behaviour of variability and light fluctuations that can be associated with stellar convection and the evolution of magnetic fields on timescales ranging from less than 1 day up to about 35 days. The light fluctuations produced by stellar activity can be described by the multifractal Hurst index that provides a measure of their persistence.}{The spot modeling indicates a lower limit to the relative surface differential rotation of $\Delta \Omega / \Omega \sim 0.02 \pm 0.01$ and  suggests a short-term cyclic variation in the starspot area with a period of $\sim 34$ days, virtually close to the synodic period of 35.2 days of the planet Kepler-30b.  By subtracting the two timeseries of the SAP and PDC Kepler pipelines, we reduce the rotational modulation and find a 23.1-day period close to the synodic period of Kepler-30c. This period also appears in the multifractal analysis as a crossover of the fluctuation functions associated with the characteristic evolutionary timescales of the active regions in Kepler-30 as confirmed by spot modelling. These procedures and methods may be greatly useful for analysing current TESS and future PLATO data.}{}
\maketitle

\section{Introduction}

Stellar rotation is a fundamental physical parameter in stellar astrophysics and plays an important role in the formation and evolution of stars \citep{kraft,sku,kawaler}. Rotation controls stellar magnetism, mixing in the stellar interior, and tidal interactions in close binary systems. Moreover, the relationship between rotation and magnetic activity  has important implications for the detectability and characterization of planets orbiting solar-like stars \citep{maxted,CollierCameron18}. Such stars comprise the vast majority of the Kepler exoplanet targets and also allow us to investigate stellar variability due to magnetic activity. In particular, starspots and active regions in the stellar photospheres modulate the stellar flux on the rotation period  \citep{walkowicz}. 

The Kepler space mission found several multi-planet systems, among which that orbiting Kepler-30 (KOI-806), that is a star with nearly solar mass and radius, an effective temperature of $5500 \pm 55$~K, and a rotation period of $\sim$16.0 days as measured by means of the Lomb--Scargle periodogram \citep{sanchis,lanza2014}. These properties make it a very interesting young solar analogue, thus motivating our choice of investigating its rotation and activity behaviour.  Moreover, it is orbited by three planets named Kepler-30b, Kepler-30c and Kepler-30d with masses of 9.2, 536 ($\sim$ 1.7 Jupiter masses) and 23.7 Earth masses, radii of 3.75, 11.98 and 8.79 Earth radii \citep[][]{panichi}, and orbital periods of 29.3, 60.3 and 143.3 days \citep[cf.][]{sanchis}, respectively. Based on the values of radius and mass, \cite{panichi} estimated that Kepler-30b is a Neptune-like planet rather than a super-Earth, Kepler-30c is a Jovian planet, while Kepler-30d is classified as a Neptune-mass planet, with bulk densities given by $\sim$0.96 g$\cdot$cm$^{-3}$, $\sim$1.71 g$\cdot$cm$^{-3}$ and $\sim$0.19 g$\cdot$cm$^{-3}$, respectively.
In this system, the orbits of the planets are on the same plane almost perpendicular to the stellar spin, a behaviour similar to our solar system. As a consequence, we observe Kepler-30 almost equator-on. This provides a fundamental constraint to reduce the degeneracies in the spot modelling of the star, that is, an additional motivation for applying our approach (see Section~\ref{spot_modelling_method}). 
By using the spot modeling, \cite{bonomo} and \cite{lanza2019} performed an analysis for the star Kepler-17, a solar-like star younger than the Sun hosting an hot Jupiter that occults starspots during transits. In both studies, the authors analysed activity and rotation of this star using starspots as tracers, comparing the longitudes of the spots mapped from the out-of-transit photometry with those of the spots occulted during transits to validate their spot modelling method. The conclusions of these studies are that spot modelling would allow us: i) to derive active longitudes where spots potentially form; ii) to estimate the lifetime of active regions; iii) to determine a lower limit for the stellar differential rotation, and; iv) to evaluate short-term and long-term activity cycles depending on the length of the available timeseries.

Inspired by the peculiar properties of the Kepler-30 system and the availability of almost four years of Kepler high-precision photometry, we decided to invest our efforts in an unprecedented analysis of this system. In particular, the present work carries on for the first time a joint analysis exploting spot modeling and multifractal methods, both widely used in the astrophysical context, but never confronted with each other.

Over the last years, several methods have been applied for an appreciation of the (multi)fractal structure of time series in astrophysical systems \citep[e.g.,][]{defreitas2016,defreitas2017,belete,francis,defreitas2019a,defreitas2019b}. The estimation of local fluctuations and long-term dependency of astrophysical time series is a problem that has been recently studied to understand the effect of long-memory processes due to the stellar rotation \citep{defreitas2019a,defreitas2019b}. As discussed by \cite{defreitas2013}, the global Hurst exponent \citep{hurst1951} is a powerful (multi)fractal indicator for analysing CoRoT and Kepler time series, which is mainly used to elucidate the persistence due to rotational modulation in the series themselves. Recently a novel technique proposed by \cite{gu2010}, known as the multifractal detrended moving average (MFDMA) method, has been applied for the multifractal characterization of Kepler timeseries \citep{defreitas2017,defreitas2019a,defreitas2019b}. The MFDMA method filters out the local trends of nonstationary series by subtracting the local means. In addition, the MFDMA method investigates the local fluctuations of timeseries using an a priori (fixed) timescale. According to \cite{wang},  fixing an a priori scaling range may lead to a crossover falling within the scaling range by mistake, and therefore the results could be biased. A method to avoid mistakes due to fixed scaling ranges is to measure the multifractal properties considering the entire timescales. To this end, a new approach named the multiscale MFDMA method (MFDMA$\tau$) will be introduced  in this paper (see Section~\ref{mfdmatau}).

Our paper is structured as follows. Observations and technical details on the Kepler mission are presented in Sect. 2. In Sect. 3, we process and prepare the Kepler-30 timeseries as obtained by the pre-search data conditioning (PDC) and simple aperture photometry (SAP) pipelines. In particular, we investigate the impact of the PDC pipeline over stellar variability when compared to the SAP data reduction, giving an overview of the characteristics of the PDC and SAP data using quality flags and discussing the difference between these pipelines. In the next sections 4, 5 and 6, we present our methods: spot modelling, classical MFDMA method, and a new approach named as multiscale MFDMA method (MFDMA$\tau$), respectively. In Sect. 7, we apply our set of methods to the Kepler-30 data and present the results of spot modelling first and then introduce the results of the application of the multifractal analysis. In the last section 8, our conclusions and final remarks are summarised.

\section{Observations}
\label{observations}
From 2009 to 2013, the Kepler mission performed 17 observational runs, a.k.a. quarters, of $\sim$90 days each, composed of long cadence  \citep[data sampling every $\sim 30$ min;][]{Jenkins2010b} and short cadence (sampling every $\sim$1 min) observations \citep{van,thompson}.
The mission public archive provides two types of data called Simple Aperture Photometry (SAP) and Pre-search Data Conditioning (PDC) time series. The SAP flux is the sum of all calibrated background-subtracted flux values of the pixels belonging to the target aperture and for which a basic calibration has been performed,  while the PDC flux is further corrected to remove instrumental and systematic effects by fitting a linear combination of the so-called Co-trending Basic Vectors (CBV) \citep{van2,kine}. These vectors describe the common trends observed in the 50\% less variable and more mutually correlated targets falling within each CCD of the focal plane \citep{Smithetal12,Stumpeetal14}.

Considering our interest in the out-of-transit variations on timescales comparable with the stellar rotation, we shall made use of the long-cadence data of Kepler-30 publicly available at the  MAST archive\footnote{\texttt{http://archive.stsci.edu/kepler}}. They covers four years and the mean relative accuracy of each data point is $\sim$260 ppm. 

\begin{figure*}
\begin{center}
\includegraphics[width=0.68\textwidth]{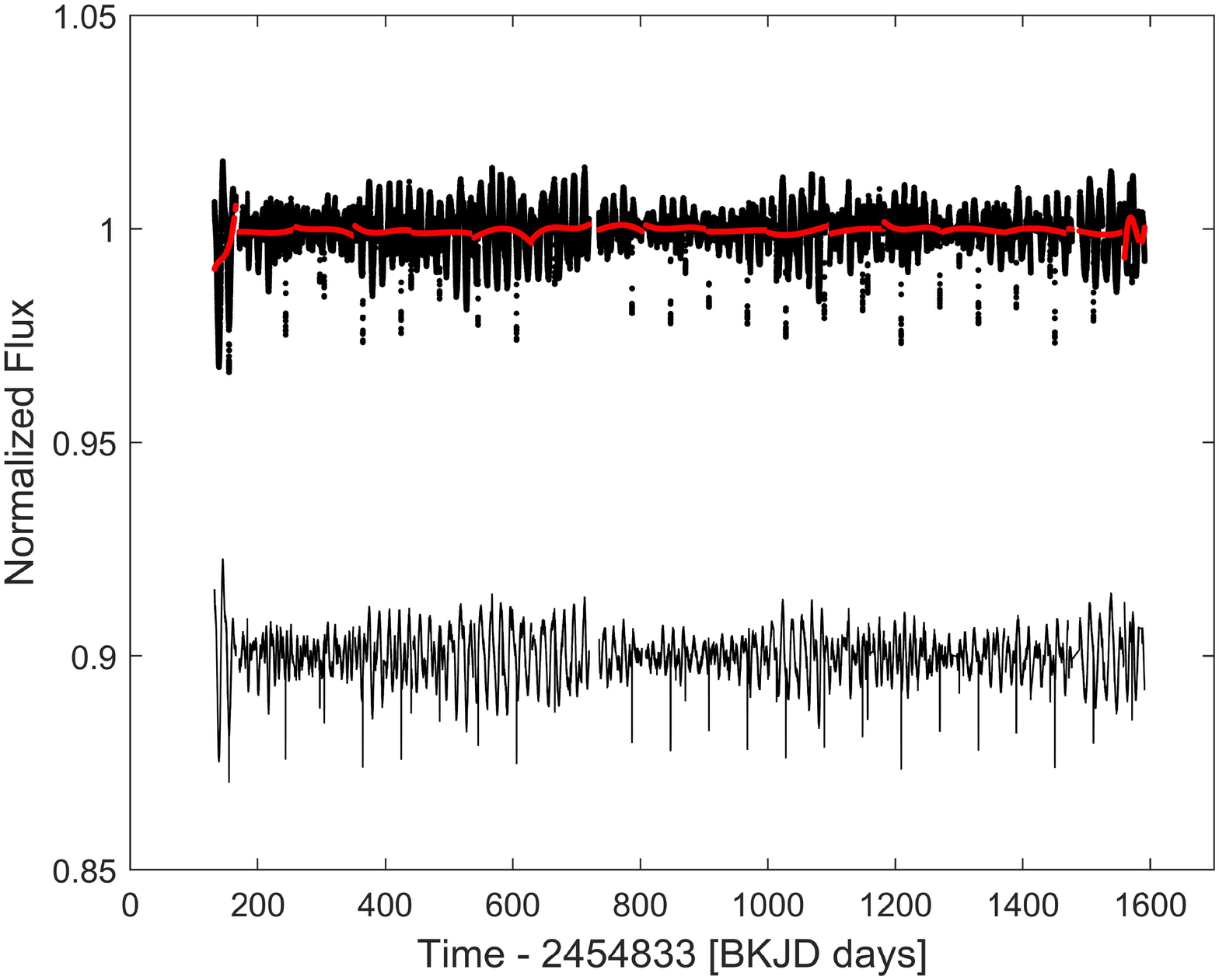}
\end{center}
\caption{Kepler-30 PDC time series. Top subfigure: PDC time series (black dots) and polynomial fits of order 3. Bottom subfigure: final PDC time series adjusted by polynomial functions. The signatures of three planets are maintained.}
\label{fig0}
\end{figure*}

\begin{figure*}
\begin{center}
\includegraphics[width=0.68\textwidth]{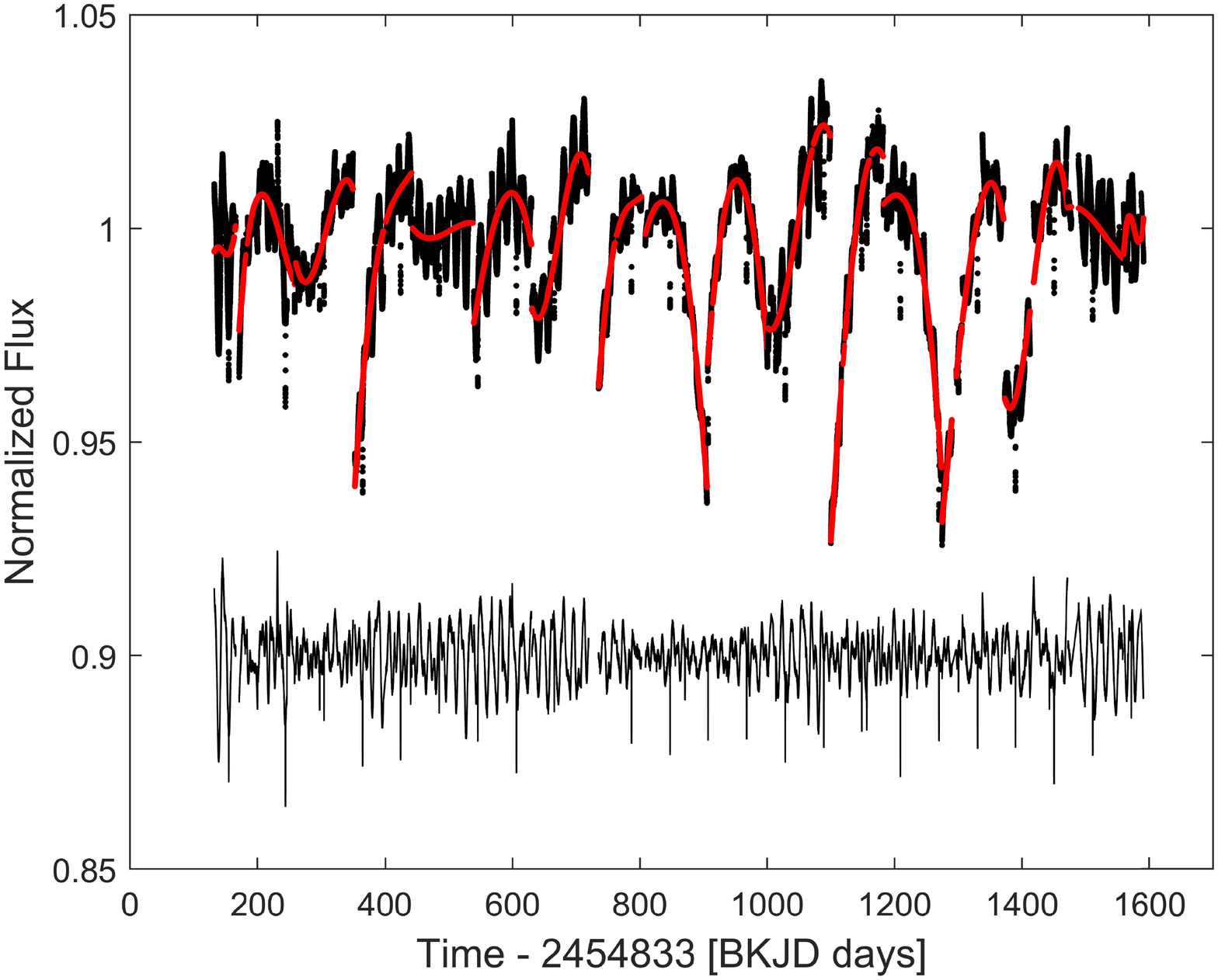}
\end{center}
\caption{Idem Figure \ref{fig0} for SAP time series.}
\label{fig0x}
\end{figure*}

\begin{figure*}
\begin{center}
\includegraphics[width=0.68\textwidth]{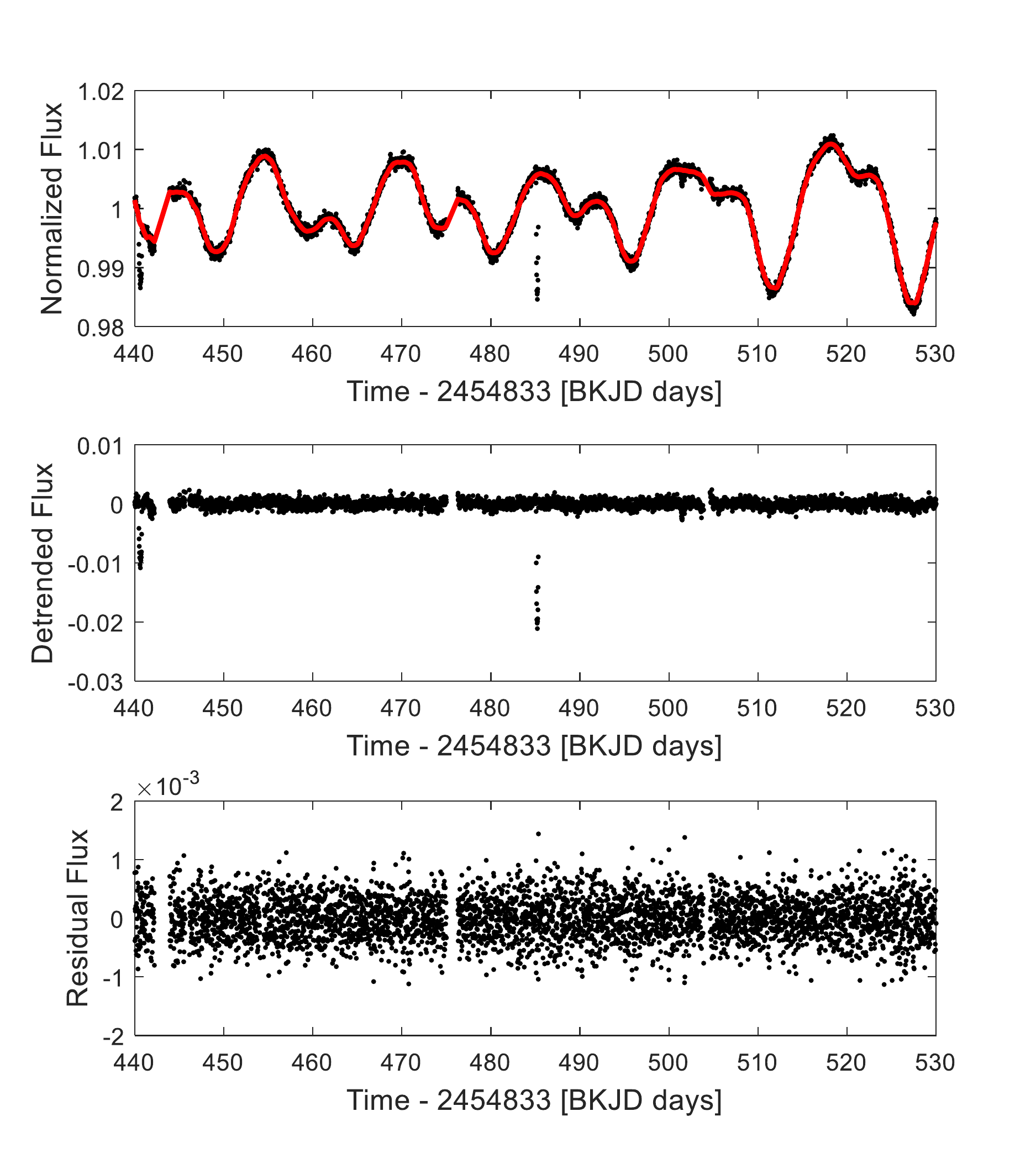}
\end{center}
\caption{Detrending of Kepler-30 with the iterative biweight filter using window size of 0.75 days (red line), which allows for a detection of the transits. Normalized flux (top), detrended flux (middle) and residual flux (bottom), both only shown for a segment of the total time series. Black dots indicate the PDC pipeline of Kepler-30 after the first three steps of our procedure. Detrended flux is a combination of noise background and planet transits, indicating the trending of the time series potentially due to rotational modulation of the star.}
\label{figLC1}
\end{figure*}

\begin{figure*}
\begin{center}
\includegraphics[width=0.68\textwidth]{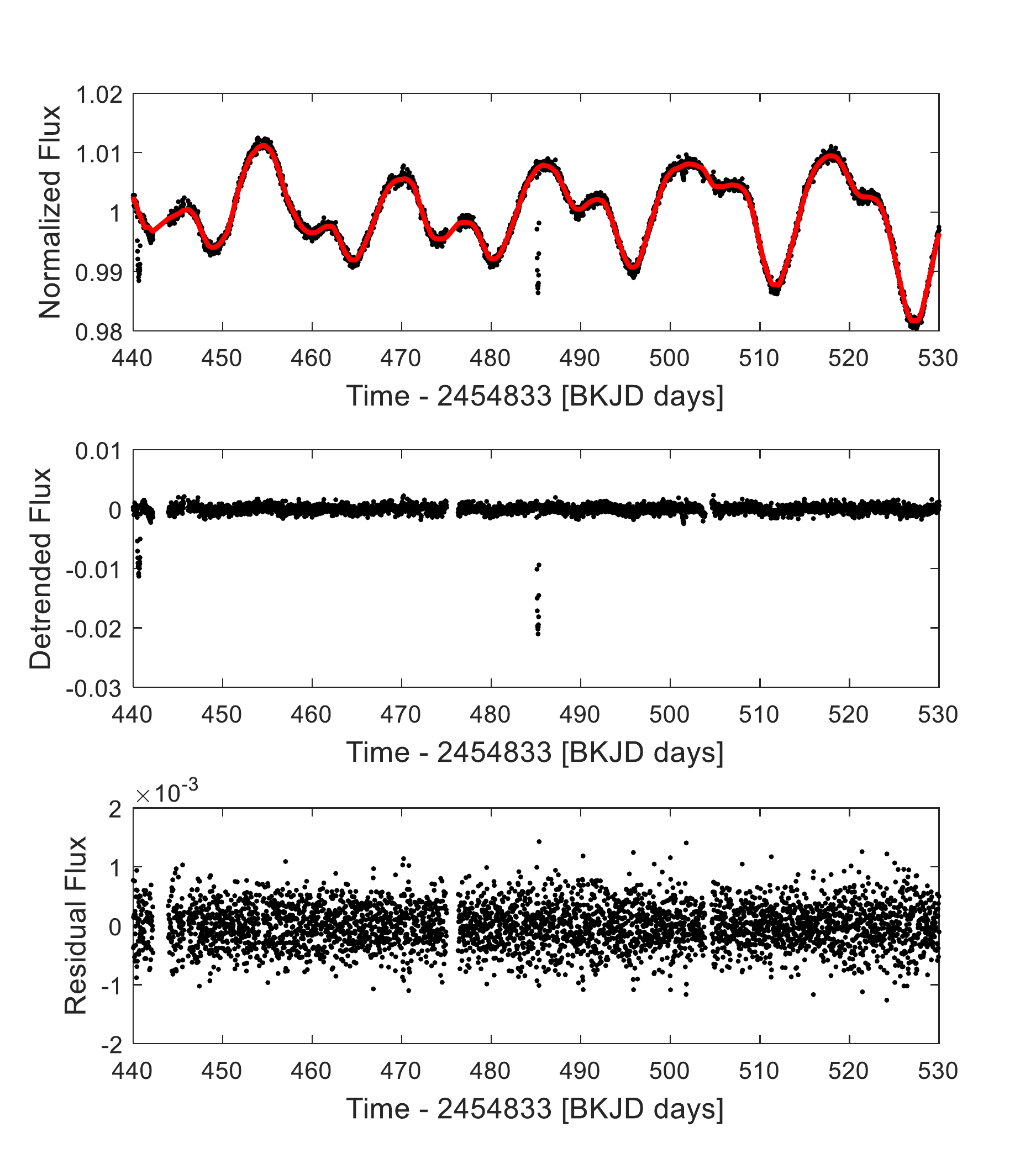}
\end{center}
\caption{Idem Figure \ref{figLC1} for SAP pipeline.}
\label{figLC2}
\end{figure*}

\begin{figure*}
\begin{center}
\includegraphics[width=0.68\textwidth]{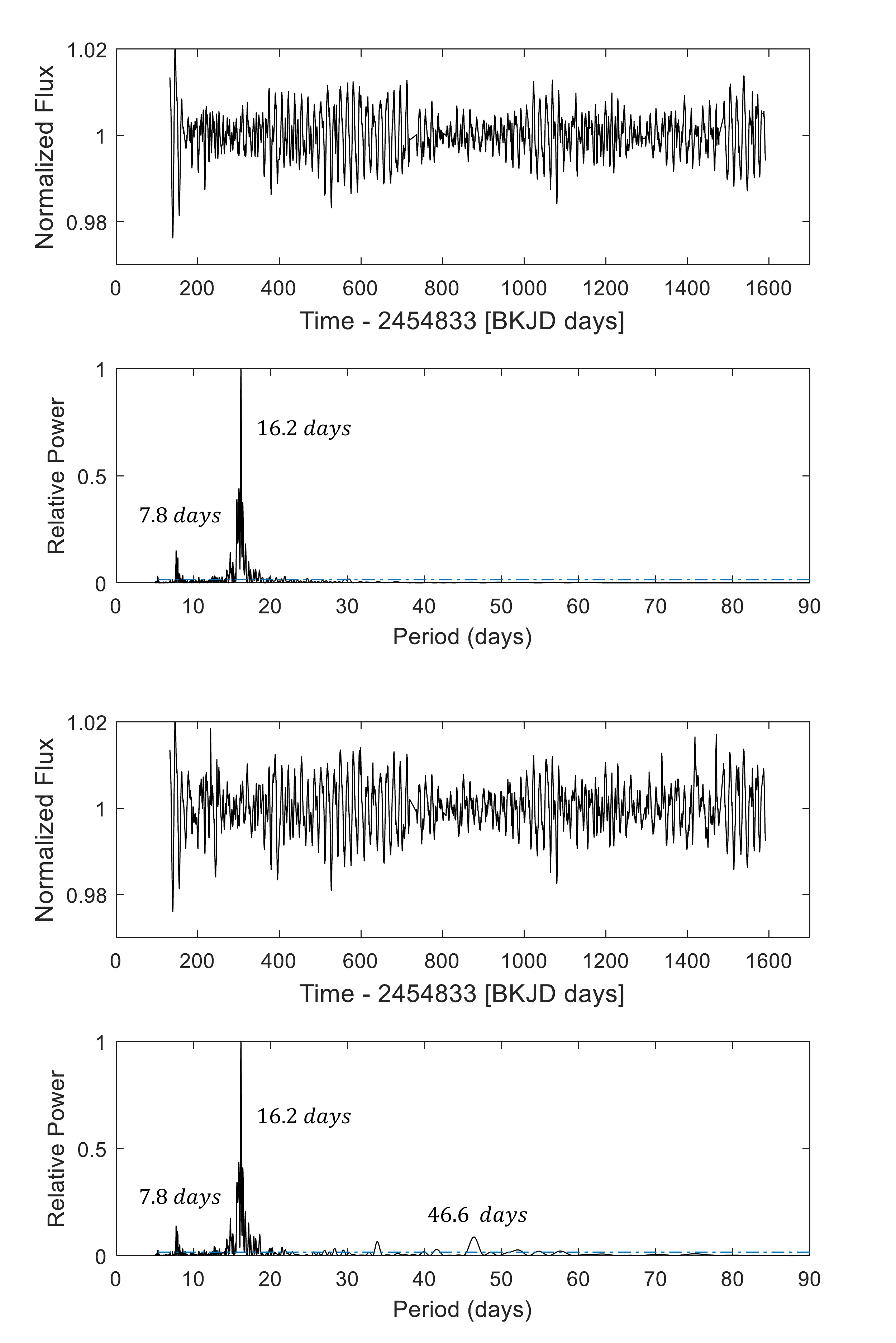}
\end{center}
\caption{Final PDC (first panel) and SAP (third panel) time series adjusted according to steps of Section 3. The Lomb-Scargle periodograms are on the bottom side correspondingly. The dashed-dot blue line is the significance threshold given by letting FAP (False Alarm Probability) equals to $10^{-4}$. For SAP data, there is a significant peak to 46.6 days also found in Figure \ref{figLS2}. As can be seen, the PDC timeseries has a flat periodogram for periods longer than 20 days confirming that the PDC pipeline tends to remove all the long-term trends.}
\label{figLS}
\end{figure*}

\begin{figure*}
\begin{center}
\includegraphics[width=0.68\textwidth]{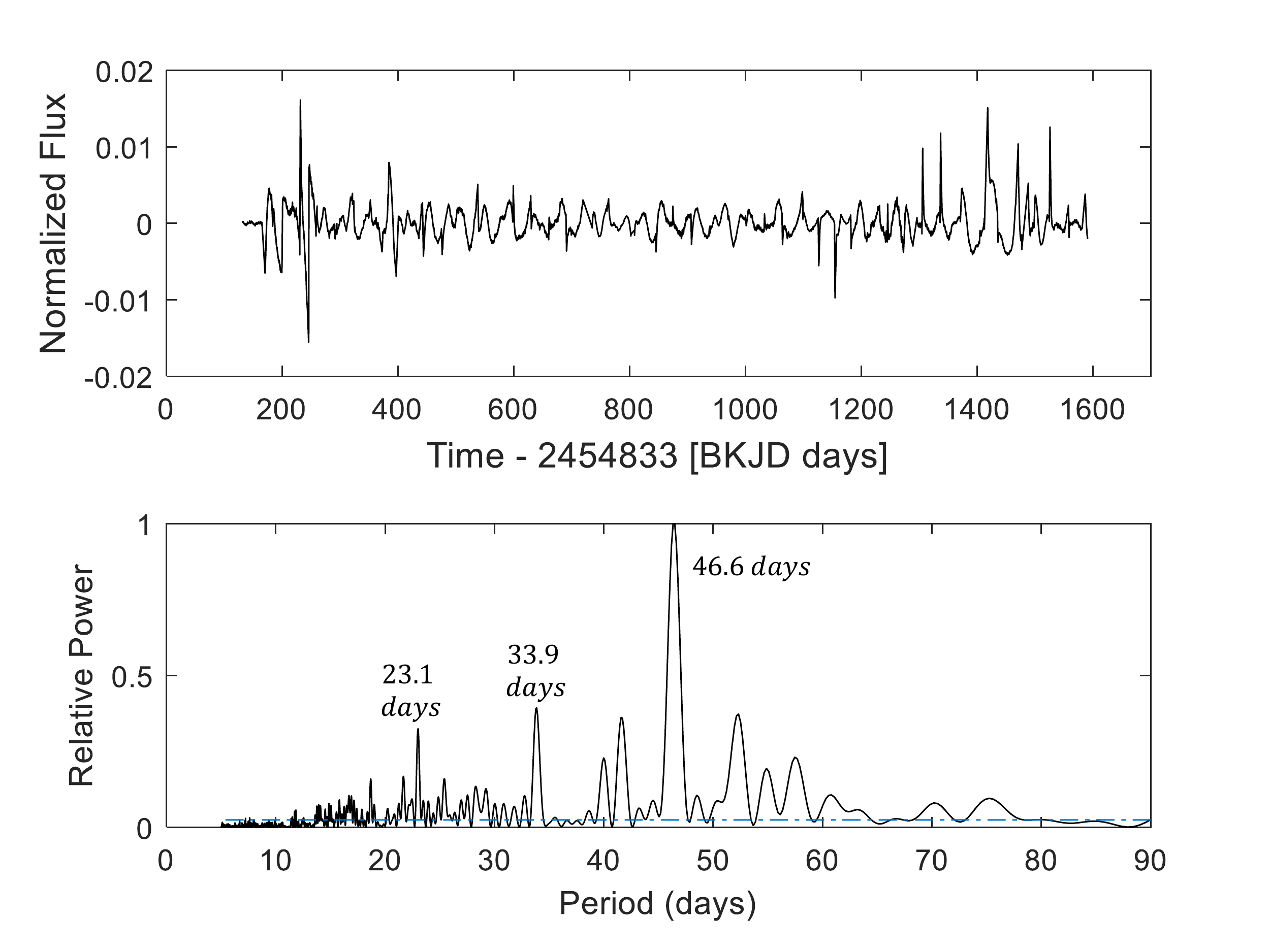}
\end{center}
\caption{Difference between the SAP and PDC time series, where the peak of 46.6 days is evidenced. A second and third peaks of 23.1 and 33.9 days are also evidenced.}
\label{figLS2}
\end{figure*}

\section{Kepler-30 preparation and pre-processing}
\label{data_preparation}
The PDC pipeline was designed to maximize the chance of detecting planetary transits and not to study stellar intrinsic variability on timescales much longer than the transit duration. Therefore, the removal of the instrumental and systematic trends by fitting a linear combination of CBVs often results in overfitting and removal of the intrinsic target variability. The effect is particularly pronounced for the brighter and more intrinsically variable targets and leads to a remarkable attenuation of the components of the variability with timescales longer than 15-20 days \citep{Stumpeetal14,gill}. This is particularly relevant for our analysis. On the other hand, the SAP timeseries, although presenting some residual instrumental effects and long-term systematic trends, is not affected by this problem. Therefore, we shall investigate the variability of Kepler-30 by analysing both the SAP and PDC timeseries. 

Because of the flux offsets between adjacent quarters and the long-term instrumental trends within each quarter \citep[cf.][]{Jenkins2010}, the time series corresponding to different quarters were separately treated as follows: 

\textbf{Firstly}, in each quarter residual outliers were removed as described in Sect. 2 of \cite{demedeiros2013}. This procedure resulted in less than 0.2$\%$ of data points being flagged as outliers and discarded. Furthermore, each Kepler timestamp has a quality flag which alerts of systematic or instrumental effects that may affect the quality of the photometric measurement. We discarded all the datapoints having the quality flag SAP\_QUALITY different from zero because they could potentially be affected by such systematics \citep{kine}. In this way, the percentage of data points removed was less than 2$\%$. 

After applying this procedure, we checked its impact by calculating the cross-correlation coefficient for zero lag (i.e., when the timeseries are not shifted) of the series before and after removing the outliers and datapoints with non-zero quality flag. To this end, we use the MATLAB function \texttt{crosscorr}\footnote{See \texttt{https://in.mathworks.com/help/econ/crosscorr.html}, for more details.}. The values of the coefficients found were 0.92 for the PDC data and 0.94 for the SAP one. The gaps in our timeseries are short enough as to have a negligible impact on our analysis. Therefore, we decided to retain them and to avoid gap filling because artifacts could arise from the necessary interpolation processes.

\textbf{Secondly}, a third-order polynomial was fitted to the data to detrend the time series, separately for each quarter, as suggested by \cite{defreitas2019b}. The planetary transits had a very small impact on the polynomial fits given the small planetary radii and their long orbital periods.  Our interest is in investigating the out-of-transit fluctuations on the timescale of stellar rotation or longer, so the deformation of the transit profiles is not an issue also in view of their removal in the next step. Figures \ref{fig0} and \ref{fig0x} show the result of the two steps described above.

\textbf{In the third step}, the \texttt{WOTAN} package proposed by \cite{hippke} was used to detrend the out-of-transit part of the signal in preparation for the transit removal in the fourth step (see below) as can be seen in the top panels of Figs. \ref{figLC1} and \ref{figLC2} (red lines). To this end, we use the iterative biweight method with a window of 0.75 days as proposed by \cite{hippke}.  
The detrended flux for both pipelines can be seen in the middle panels of Figs. \ref{figLC1} and \ref{figLC2}. 

\textbf{In the fourth step}, planetary transits were removed according to the parameters given in Table 1 of \cite{sanchis}. We also performed a visual inspection and checked that the obtained out-of-transit series has no evident transit signal left after the application of our procedure. The gaps introduced by discarding the transits do not exceed 9 hours, i.e., the duration of the transit of the most distant planet Kepler-30d. These gaps shall have a negligible effect on our spot modelling because the rotation period of the star is $\sim$16 days. In the same way as mentioned above, we did not perform any interpolation process to fill in gaps, since the number of data removed is negligible. Examples of the final results of our procedure can be seen in the bottom panels of Figs. \ref{figLC1} and \ref{figLC2}.

\textbf{Lastly}, the final time series is obtained by combining the 17 quarters and consists of four years of data. From each quarter series,  the median value was subtracted and then all the residual quarter series were stitched together. The final timeseries are shown in the upper panels of Figs. \ref{figLS} and \ref{figLS2}, where the median of the errors of the single photometric measurements is 2$\times 10^{-4}$ in relative flux units. The number of datapoints is $N=60007$ and $N=60018$ for the PDC and SAP timeseries, respectively. 

\subsection{Lomb--Scargle periodograms and residual flux between SAP and PDC pipelines}

The final PDC and SAP time series and their Lomb--Scargle periodograms \citep[cf.][]{lomb,scargle} with the significance levels are illustrated in Fig. \ref{figLS}. For both timeseries, there are two main peaks at 7.8 and 16.8 days. Nevertheless, for SAP time series, a lower peak at 46.6 days also appears. As mentioned above, the SAP pipeline preserves other  variabilities that are attenuated or removed by the PDC pipeline. Consequently, the Residual Time Series (hereafter RTS) (i.e., the difference) between the SAP and PDC time series can reveal the origin of that peak (see upper panel of Fig. \ref{figLS2}).

In Fig. \ref{figLS2} (bottom panel), we plot the Lomb--Scargle periodogram of the difference time series (RTS). Only a few points were lost during the process of subtracting the time series.  The Lomb--Scargle periodogram of Fig. \ref{figLS2} clearly shows that the period of 46.6 days, as well as the periods of 23.1 and 33.9 days, are preserved by the SAP pipeline. The Lomb--Scargle technique has a clear disadvantage, because it is limited to analysing in the frequency space and therefore is not capable of resolving the time-evolution of a given periodicity. In addition, this technique does not allow us to analyse the behaviour of the variability, and of the light fluctuations in particular, over different timescales and amplitudes. To this end, we shall consider now the spot modeling (cf. Sect.~\ref{spot_modelling_method}) and the MFDMA approaches (cf. Sects.~\ref{mf_gen_descr}and~\ref{mfdmatau}). 

\section{Spot modeling of Kepler-30 time series}
\label{spot_modelling_method}
The small $v\sin i \sim 1.94 \pm 0.25$ km~s$^{-1}$ of Kepler-30 \citep{fabrycky} prevents the application of Doppler Imaging techniques to this star \citep{Strassmeier09a}, therefore, the only possibility of mapping its surface is a spot modeling based on the inversion of its light curve. 

Spot modeling is notoriously an ill-posed problem because it tries to extract a bi-dimensional map from a one-dimension timeseries. To account for the inherent degeneracies of the solution, we apply a regularization to our inversion. It consists in introducing a priori information to select the unique spot map that maximizes the configurational entropy of the spot pattern for a given level of the chi square of the fit. We choose the configurational entropy to be maximal when the star is unspotted, that is, when the spot map has the least spotted area.

The fundamental assumption of our spot model is that the spot pattern stays stable for a time interval $\Delta t_{\rm f}$ during which the flux received from the star is modulated only by stellar rotation. A moderate random starspot evolution on a timescale significantly shorter than $\Delta t_{\rm f}$ is not a problem because it contributes to the noise level and does not affect the spot modeling thanks to the application of the Maximum Entropy (hereinafter ME) regularization (see Appendix~\ref{appendix_spot_model}). However, sizeable changes in the spot pattern on a timescale comparable with $\Delta t_{\rm f}$ can produce systematic errors in the spot map. An optimal selection of $\Delta t_{\rm f}$ is therefore relevant for our modeling and is discussed in Appendix~\ref{appendix_spot_model} together with the determination of the other parameters of our model. 

A main advantage of modeling a star with transiting planets such as Kepler-30 is the possibility of measuring the obliquity of its spin axis projected on the plane of the sky that allows us to constrain the inclination $i$ of the stellar spin to the line of sight \citep{sanchis}. Given the low projected obliquity,  we assume the stellar spin to be normal to the plane of the orbit of the more massive close-by transiting planet Kepler-30c, that is, virtually perpendicular to the line of sight, thus fixing the geometry of the rotating star to be mapped (see Appendix~\ref{appendix_spot_model} for the adopted value of the inclination). Note that in the case of a star with an unknown inclination, spot modeling is much more uncertain because of the  strong degeneracy between the inclination and the latitudes of the spots.  

Since we observe Kepler-30 virtually equator-on, spot modeling is unable to provide information on the latitude of the starspots because the duration of their visibility along a stellar rotation is independent of their latitude (we neglect the effect of latitudinal differential rotation the value of which is unknown a priori). Therefore, we can only map the distribution of the spotted area vs. the longitude and the time along successive intervals of duration $\Delta t_{\rm f}$. As a consequence, we consider two-dimensional (2D) spot maps only as an intermediate step in our modeling and derive the distribution of the spotted area vs. longitude from them,  that is,  compute  one-dimensional maps of the filling factor by integrating over the latitude in the 2D maps. Such longitude distributions of the filling factor can be directly related to the light modulation under the assumption that the spot contrast is constant. This greatly reduces the degeneracies associated with the non-uniqueness of light curve inversion.

We include the effect  of solar-like faculae in our model, but their contribution to the flux modulation is found to be very small and can be neglected for a star as active as Kepler-30 where dark spots dominate \citep[cf.][see also Appendix~\ref{appendix_spot_model}]{Radicketal18}. 
More details of our spot modeling and the impact of the modeling parameters on the spot maps are presented in Appendix~\ref{appendix_spot_model}. 

Recent works have questioned the validity of spot models assuming a few discrete spots on the stellar surface \citep[e.g.,][]{BasriShah20}. Our model is based on a continuous distributions of spots and solar-like faculae on the stellar photosphere and has been extensively tested in the case of the Sun \citep{Lanzaetal07} as well as of active stars with transiting planets by comparing the spot maps from light curve inversions with the real distributions of the sunspot groups and the spots detected by occultation during planetary transits, respectively \citep{Silva-ValioLanza11,lanza2019}. Thanks to such comparisons, we are confident that our spot modelling is capable of reconstructing the distribution of the spot area vs. the longitude with a resolution of $\approx 60^{\circ}$ in the case of Kepler-30. In consideration of the systematic differences introduced by the Kepler pipeline between the SAP and the PDC timeseries, we shall apply our modelling to both of them and compare the results to look for artifacts produced by those systematics.

\section{Multifractal description of timeseries}
\label{mf_gen_descr}
\subsection{Basic concepts and definitions}
The light variations in the time series of Kepler-30 display a wide range of timescales ranging from tens of minutes characteristic of photospheric convection,  to the rotational modulation due to photospheric brightness inhomogeneities (starspots), up to months or years characteristic of stellar activity cycles. The rotational modulation signal is quasi-periodic and non-stationary owing to the evolution of starspots on the surface of the star. To characterize the fluctuations on the shortest timescales, mostly produced by stellar convection, we can use methods of multifractal timeseries analysis, that we briefly introduce below. We mainly follow the treatment by \cite{Kantelhardt15} to whom we refer the reader for a more detailed account because here we limit ourselves to a minimal introduction with the purpose of making this paper  accessible also to readers who do not possess previous knowledge of the field. 

To characterize the wide frequency spectrum of convective fluctuations, it is useful to look for self-similarity properties in the fluctuations themselves. Specifically, we define $\Delta f (t)$  as the light fluctuation at the time $t$, measured with respect to some constant mean value. A rescaling of the time by an arbitrary factor $a$, i.e., $t \rightarrow a t$ may require a re-scaling of the fluctuations by a factor $a^{H}$, that is, $\Delta f(t) \rightarrow a^{H} \Delta f(at)$, in order for the re-scaled fluctuations to follow the same statistical distribution as the initial ones. A timeseries that obeys this scaling property for an arbitrary factor $a$ is called a self-affine or self-similar timeseries and the exponent $H$ is called the Hurst exponent of the timeseries. Several stochastic fluctuations, including those associated with turbulent convection, satisfy this definition.  

The Hurst exponent $H$ characterizes the kind of self-similarity. The simplest example is the timeseries of the mean-square displacement $x^{2}(t)$  of a point performing a random walk with a normally distributed step size. In this case, the time scaling $t \rightarrow a t$ requires a corresponding scaling $ x \rightarrow a^{1/2} x$ in order to preserve the statistical distribution of the variable $x$. In other words, the timeseries  $x(t)$ exhibits a self-similar appearance with $H=0.5$, that is, by expanding the time axis by a factor $a$ and the space axis by a factor $a^{H}$ the distribution of the statistical fluctuations of $x(t)$ is the same as the original one. Such a scale self-similarity (or scale invariance) is one of the fundamental properties that define fractals. 

Self-affine timeseries are persistent in the sense that a large fluctuation is likely followed by another large fluctuation and a small fluctuation  by a small one. The persistence holds on all the timescales for which the self-affine scaling holds. In more complex systems, the persistence may change according to the timescale being stronger on some timescales and weaker on others. For example, the weather timeseries are usually persistent on timescales of hours or days, but the degree of persistence is higher on shorter timescales. Moreover, the characteristic persistence timescale generally changes with the season usually being longer during winter and summer and shorter in autumn and spring. 

Considering a stochastic self-affine timeseries $x(t)$ with the variable sampled at equidistant times $t_{i}$, with $i=1, 2, ..., N$, we define the timeseries of the increments $\Delta x_{i} \equiv x(t_{i}) - x(t_{i-1})$. In the case of the simple random walk model considered above with $H=0.5$, the $\Delta x_{i}$ are independent of each other, but when $H> 0.5$, a positive $\Delta x_{i}$ is likely to be followed by a positive increment (persistence), while, when $H< 0.5$, by a negative one (anti-persistence). The degree of persistence can be quantified by the auto-correlation function in the case of a stationary timeseries, that is, one with constant mean and standard deviation. If we denote the timescale as $n$, also indicating a number of datapoints in a uniformly sampled timeseries, the autocorrelation is defined as:
\begin{equation}
C(n) = \frac{1}{\sigma^{2}} \frac{1}{N-n} \sum_{i=1}^{N-n} \Delta x_{i} \Delta x_{i+n},
\end{equation}
where $\sigma^{2}$ is the variance of the timeseries of the increments $\Delta x_{i}$. 

When the $\Delta x_{i}$ are uncorrelated, $C(n) = 0$ for $n > 0$. A short-range correlation can be described by an exponentially decaying $C(n) \propto \exp( -n/n_{\rm d})$, where $n_{\rm d}$ is the decay timescale. For example, this is the case of an autoregressive process defined by the linear recursion $\Delta x_{i} = c \Delta x_{i+1} + \epsilon_{i}$, where $c = \exp(-1/n_{\rm d})$ and the $\epsilon_{i}$ are normally distributed random deviates. 

In the case of long-term correlations, the asymptotic behaviour of $C(n)$, that is, its scaling for large $n$, can be described by a power law: 
\begin{equation}
C(n) \propto n^{-\gamma},
\end{equation}
where the exponent $0<\gamma <1$ characterizes the long-term correlations. It can be shown that a long-term correlated, i.e., persistent, behaviour of $\Delta x_{i}$ leads to a self-affine scaling of the $x(t_{i})$ characterized by an Hurst exponent $H = 1- \gamma/2$. Therefore, the Hurst exponent can be used to characterize the autocorrelation function of the increments of the timeseries. 

Ideally, the estimation of the exponent $\gamma$ can be made by computing the autocorrelation function of the increments or the power spectrum of the original time series $x(t_{i})$ \citep{Kantelhardt15}. In practice, these approaches are doomed to fail because of the noise that dominates the autocorrelation and the power spectrum in the asymptotic regime of large $n$ values  to be considered when deriving $\gamma$. Therefore, more robust statistical approaches have been developed for such a determination. They are based on the so-called {\em fluctuation functions} $F_{q} (n)$ that measure the scaling of the fluctuation amplitude with the timescale $n$, where $q$ is a parameter defining the moment of the statistical distribution of the $x(t_{i})$ measured by the function itself (see Appendix~\ref{multifractal_background} for examples of fluctuation functions). The fluctuation function based on the standard deviation corresponds to $q=2$ and plays a special role in the analysis. 

The scaling of the fluctuation function with $n$ is characterized by a generalized Hurst exponent or Holder exponent $h(q)$ that depends on the moment of the distribution sampled by $F_{q} (n)$. To define the fluctuation function $ F_{q} (n)$, we start from  the root-mean-square (RMS) fluctuation function $F_{\nu} (n)$ that is defined by
\begin{equation}
F_{\nu} (n) = \left\{ \frac{1}{n} \sum_{i=1}^{n} \epsilon_{\nu}^{2} (i)\right\}^{\frac{1}{2}},
\end{equation}
where  $\epsilon_{\nu}(i)$ is the deviation of the $i$-th datapoint from a suitable moving average introduced to remove the long-term variation in the timeseries (see Appendix~\ref{multifractal_background} for its precise definition). Starting from the RMS fluctuation function, we define the fluctuation function of order $q$ as 
\begin{equation}\label{fluctu}
F_{q}(n)=\left\lbrace \frac{1}{N_n} \sum_{\nu=1}^{N_n} F^q_\nu(n) \right\rbrace^{1/q} \textrm{for}\; q\neq0
\end{equation}
and, for $q=0$,
\begin{equation}\label{fluctuzero}
\ln\left[F_{0}(n)\right]=\frac{1}{N_{n}}\sum^{N_{n}}_{\nu=1}\ln [F_{\nu}(n)],
\end{equation}
where $N_n$ is the number of  non-overlapping segments  for a given segment of size $n$ in the timeseries, i.e., $N_{n} \sim N/n$, where $N$ is the total number of datapoints in the series (see Appendix~\ref{multifractal_background} for the precise definition of $N_{n}$). For larger values of $n$, the fluctuation function scales following a power-law given by 
\begin{equation} \label{Fqxn}
F_{q}(n) \sim n^{h(q)},
\end{equation}
that allows us to define the generalized Hurst exponent $h(q)$. 

The previously defined Hurst exponent corresponds to the value attained for $q=2$, that is, $H \equiv h(2)$. By definition, when $h(q)$ is constant, we speak of a fractal (or monofractal) time series, while when $h$ depends on $q$ we have multifractal time series. In the former case, the scaling exponent of the fluctuation self-affine law is independent of the amplitude of the fluctuations, while in the latter case it depends on their amplitude because different values of $q$ corresponds to different moments that sample fluctuations of large or small amplitudes in different ways (cf. equation~\ref{eq4}).

The scaling of the fluctuations with their amplitude can be quantitatively characterized by means of the multifractal spectrum and its characteristic parameters as detailed in Appendix~\ref{multifractal_background}. Multifractal timeseries can be produced, for example, by simple, non-linear, recurrence laws, such as the so-called logistic law $x_{i+1} = \lambda_{\rm ll} x_{i} (1-x_{i})$, where the coefficient $\lambda_{\rm ll}$ falls in a range leading to a chaotic behaviour \citep[e.g.,][]{LuoHan92}.

\subsection{Origins of multifractality and the impact of quasi-periodic modulations in the timeseries}
\label{origins_of_mf}

The multifractal character of a stochastic timeseries is manifested by different scaling exponents for small and large fluctuations on the same timescale $n$. This behaviour can be originated by the simultaneous presence of different long-term correlations and persistences that depend on the amplitude of the fluctuations and/or by their statistical distribution that, even in the absence of long-term correlations, deviates strongly  from a Gaussian (normal) distribution. In the latter case, the distribution of the fluctuations usually shows fat tails as, for example, in the case of a Cauchy distribution. 

It is possible to discriminate between these two sources of multifractality by considering the so-called surrogates of the original timeseries. A first kind of surrogate is simply produced by randomly shuffling the datapoints over the set of the $N$ indexes $i$. In this way, we keep the statistical distribution of the datapoints, but remove all short- and long-term correlations producing a new timeseries with $h(q)$ close to 0.5, that is the Hurst exponent of a random time series consisting of uncorrelated points. A second kind of surrogate can be obtained by taking the Fourier transform of the original timeseries and constructing a modified Fourier transform having the same modulus, but a phase randomly extracted in the $[-\pi, \pi]$ interval. The backward transformation of this modified transform into the time domain gives a timeseries the points of which retain most of the correlations present in the original timeseries, but whose distribution is nearly Gaussian, thus effectively eliminating the effect of non-Gaussianity in the original distribution \citep[e.g.,][]{Kantelhardt15,wang}. 

By comparing the $h(q)$ function and the multifractal spectrum of the original timeseries with those of the above surrogates, we can extract information on the dominant source of multifractality. Specifically, when the dominant source of multifractality is time correlation, the $h(q)$ function of the randomly shuffled series will be remarkably different from that of the original series, while the $h(q)$ of the phase-randomized series will be similar. On the other hand, when the dominant source of multifractality is a non-Gaussian statistical distribution of the datapoints, the converse will be true. 

From a more general point of view, timeseries can result from both deterministic and stochastic processes, the former usually being responsible for long-term trends or oscillations.  Those trends should be removed to allow the best determination of the properties of the stochastic fluctuations. This is usually done  by fitting the trends with polynomials of different degrees in the calculation of the fluctuation functions \citep[e.g.][]{Kantelhardt15} or by subtracting a moving average as in our case (see Appendix~\ref{multifractal_background}). However, in the case of astrophysical systems, and of magnetically active stars in particular, a clear separation between deterministic long-term trends and short-term stochastic fluctuations is not always possible. For example, the rotational modulation of the flux due to photospheric starspots contains a stochastic component produced by the evolution of the individual starspots often appearing randomly in time and/or longitude \citep[e.g.][and references therein]{lanza2019}. 

In light of such an impossibility of performing a clear separation between the different components of the photometric timeseries of magnetically active stars, the adopted approach has been that of including the rotational modulation and the evolution of the surface brightness inhomogeneities as additional sources of multifractality. In other words, the multifractal analysis itself has been used as a  technique to characterize those processes in addition to standard techniques such as Lomb-Scargle periodogram or autocorrelation to measure stellar rotation periods \citep[e.g.,][]{McQuillanetal13} or spot modelling to derive the evolution of the surface features \citep{Lanza16}. 
This approach has been extended even to the detection and characterization of strictly periodic phenomena such as planetary transits in the presence of noise \citep[e.g.,][]{Agarwaletal17}. 

In a sample of magnetically-active solar-like stars observed by the CoRoT space telescope, \cite{defreitas2013} found a correlation between the Hurst index and their rotation period by assuming that their timeseries were monofractal. In a subsequent work on the same sample, \cite{defreitas2016} established that their timeseries are indeed multifractal, that is, the correlations produced by magnetic activity and rotation are better described by assuming a generalized Hurst exponent $h(q)$ and adopting the full set of parameters introduced in Appendix~\ref{multifractal_background} to characterize the multifractal spectrum, in particular $\Delta \alpha$. 

Later, \cite{defreitas2017} analysed a sample of 34 M dwarf stars previously investigated by \cite{mathur2014a}, to test the behaviour of the Hurst exponent against the magnetic activity indicator $S_{\rm ph}$. The latter index is based on the standard deviation of the flux in the Kepler passband computed over non-overlapping time intervals of $k$ mean rotation periods $P_{\rm rot}$ after removing the contribution of the photon shot noise to the standard deviation. An average index $\langle S_{\rm ph} \rangle_{k}$ is then defined as the arithmetic mean of all the individual $S_{\rm ph}$ indexes determined from each of the $k P_{\rm rot}$ intervals along the timeseries. Therefore, it includes the contributions of all the sources of light variations from the shortest timescales up to $kP_{\rm rot}$, including the rotational modulation. On the other hand, the index $ H = h (2)$ is defined as the scaling exponent of the RMS fluctuation function  $F_{2} (n) \propto F_{\nu} (n) \sim n^{H}$ that is evaluated with the fluctuations obtained after subtracting a moving average from the timeseries. This gives a relatively greater weight to the  shorter timescales, while reducing the contribution of the rotational modulation and longer timescales. Therefore, with a generally adopted value of $k=5$ \citep{mathur2014a}, the correlation between the $H$ index and the $\langle S_{\rm ph} \rangle_{k} $ index is expected to be low because of such a difference in  the respectively dominant timescales. This expectation was  confirmed by \cite{defreitas2017} who found a slight anti-correlation between the two indexes with a Pearson  coefficient of  $-0.33$ when considering the above mentioned sample of M dwarf stars (see their Figure~6). 

On the other hand, even within a limited range in the rotation period ($<15$ days), \cite{defreitas2017}  confirmed the strong correlation between $H$ and $P_ {\rm rot}$ and found the Hurst exponent to be an indicator of magnetic activity. Recently, \cite{defreitas2019a,defreitas2019b} extended the field of application to timeseries showing more than one significant rotational periodicity, which can be considered as an indication of stellar differential rotation. Using a large sample of Kepler active stars, \cite{defreitas2019a} showed that in the relationship between the Hurst exponent and the relative amplitude of the differential rotation $\Delta P/P$ a strong trend of increasing $\Delta P/P$ toward larger $H$-index is observed. In addition, the study shows that the correlation is stronger for the most active stars in the sample.

\section{MFDMA analysis}\label{mfdmatau}

Recently, \citet{defreitas2019a} used multifractal analysis to investigate the multi-scale behaviour of a set of $\sim$ 8000 active stars that were observed by the Kepler mission. \cite{defreitas2016,defreitas2017,defreitas2019a,defreitas2019b} showed that the multifractal detrending moving average (MFDMA) algorithm, which was developed by \cite{gu2010}\footnote{MATLAB codes for MFDMA analysis can be found in the \texttt{arXiv} version of \cite{gu2010}'s paper: \texttt{https://arxiv.org/pdf/1005.0877v2.pdf}} and \cite{tang}, is a powerful technique that provides valuable information on the fluctuations of a time series. A complete description of the MFDMA algorithm can be found in Appendix A. 

Several authors \citep[e.g.,][]{gier,wang} have suggested that a single scaling exponent ($H=h(2)$) is inadequate to describe the internal dynamics of complex signals, and consequently, classical (multi)fractal methods (e.g., DFA and MFDFA) have been modified for an estimation of the temporal dependence of the spectrum of scale exponents. These improved methods are used to better quantify the short- and long-range correlations, that is important in our case because multifractality in our timeseries arises mostly because of the correlations (cf. Sect. 4.2 and Table \ref{tab1}). 

Inspired by these studies, we decided to introduce another method to calculate a time-dependent scaling exponent $h(q, \tau)$, where $\tau$ is a timescale (see below).
We call it the multiscale MFDMA method (MFDMA$\tau$) and it allows us to extend the investigation of stellar variability by including a time dependence in the description of the multiscale fluctuations. 
MFDMA$\tau$ is relatively immune to additive noise and non-stationarity, including the non-stationarity due to occurrence of events of a different dynamics.

To explore the multiscale fluctuations in flux, we introduce the increments $\Delta x(t,\tau)=x(t+\tau)-x(t)$, where $\tau$ varies from 29.4 min (Kepler cadence) to $\sim$60 days. We apply the MFDMA analysis to the increment time series $\Delta x(t, \tau)$, thus obtaining an Hurst exponent $h(q,\tau)$ that is a function also of the timescale $\tau$ and in this way introduce the MFDMA$\tau$ method. The consideration of the increment timeseries in the MFDMA$\tau$ approach strongly reduces the effects of all the stationary modulations having a period equal to $\tau$ or to one of its multiples. In other words, this method allows us to isolate the effects of the fluctuations after removing those periodicities. Given the remarkable rotational modulation present in our time series, this method will allow us to study the properties of the fluctuations after removing the modulation itself with an appropriate selection of the timescale $\tau$.  

\subsection{Optimal delay time $\tau$}
\label{tau}

\begin{figure*}
\begin{center}
\includegraphics[width=0.68\textwidth]{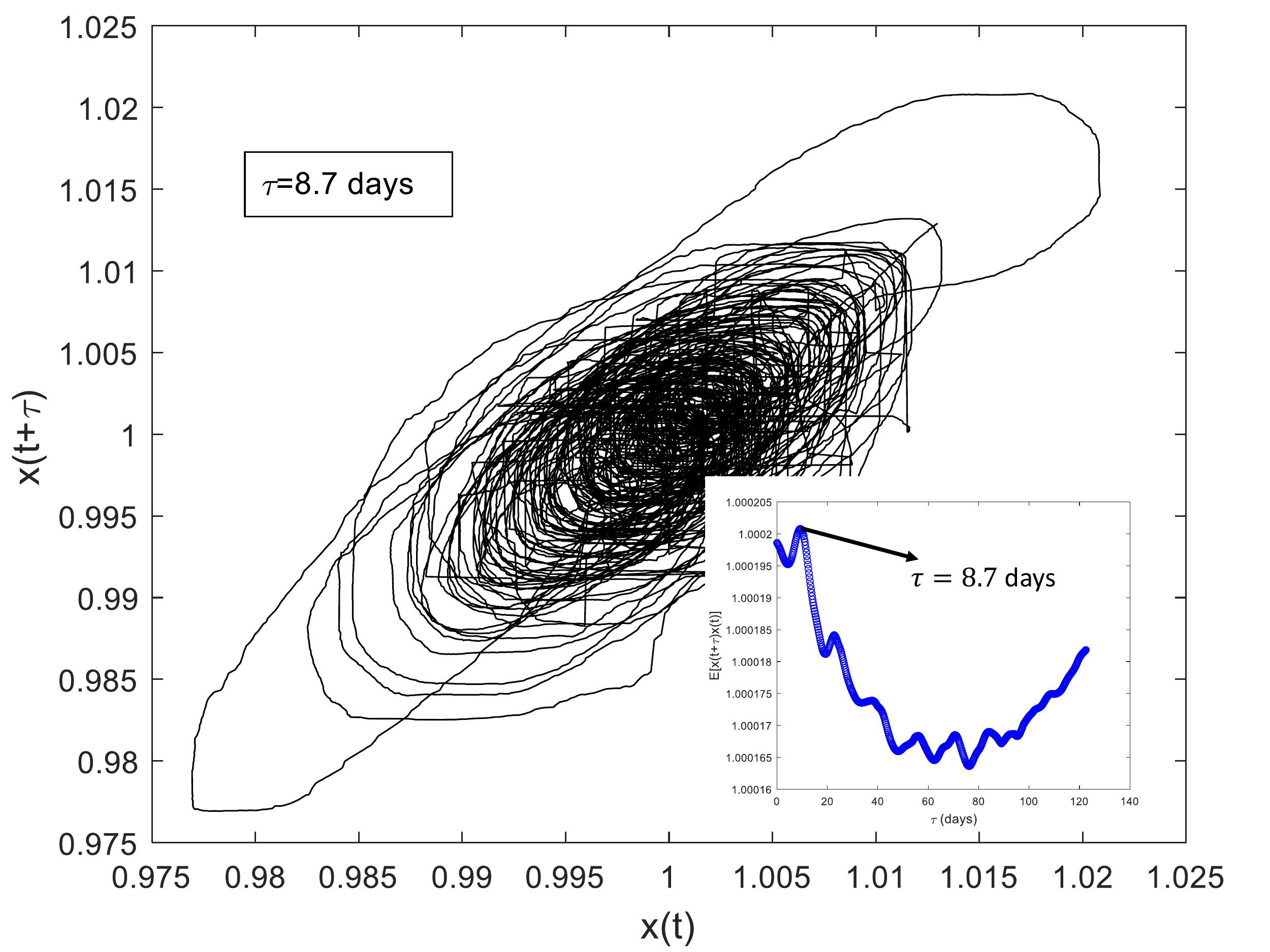}
\end{center}
\caption{The time-delayed PDC timeseries $x(t+\tau)$ with the optimal delay $\tau=8.7$~days vs. the original time series $x(t)$. The subplot shows that $\tau=8.7$ days corresponds to the maximum value of the autocorrelation function of the timeseries.}
\label{figFirst3}
\end{figure*}

We explored different values of $\tau$ using a plot of $x(t+\tau)$ vs. $x(t)$ following an approach reminiscent of that used to study multifractal attractors by means of plots of the trajectory of the dynamical system in a bi-dimensional phase space.  In Fig.~\ref{figFirst3} we plot the case corresponding to the optimal value $\tau=8.7$ days that shows a repetitive, quasi-stationary pattern organized around a fixed point with a behaviour similar to that of an attractor.

According to \cite{henry}, the optimal delay time $\tau$ can be determined using the approach of  topological properties based on the behaviour of the autocorrelation function. There are various proposals for choosing an optimal delay time using the autocorrelation function. We choose the best value for $\tau$ as that corresponding to the maximum of the mean value of the autocorrelation coefficient. 
In the subplot of Fig. \ref{figFirst3}, we show that the mean autocorrelation coefficient of the stochastic process $x(t)$, defined as $E[x(t+\tau)x(t)]$ (for all $t$), is maximum for $\tau=8.7$ days. 

\section{Results and Discussions}
\label{results}
We present the results of our spot modelling first and then introduce the results of the application of the multifractal analysis to the original photometric timeseries of Kepler-30 as well as to the residuals of the spot models. In this way, we can apply the results of the spot modelling to elucidate those of the multifractal analysis, discussing the characteristic rotational and activity timescales found in the data. 

\subsection{Best fit of the Kepler-30 light curves}
\label{best_fit}
\begin{figure*}
\begin{center}
\includegraphics[width=0.88\textwidth]{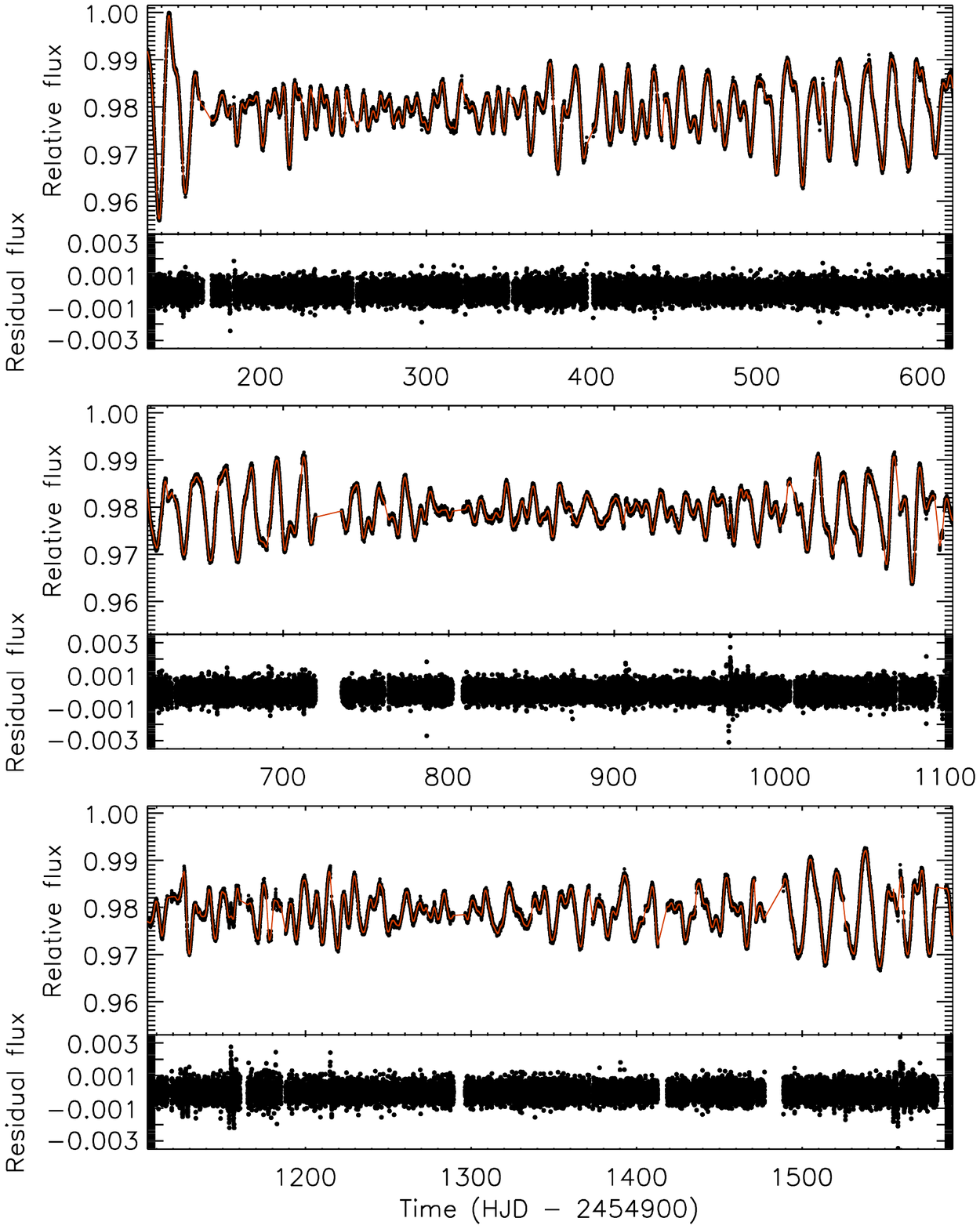}
\end{center}
\caption{Upper panels: The PDC light curve of Kepler-30 (solid black dots) and the unregularized best fit obtained with our spot model (red solid line). Lower panels: the residuals of the best fit. }
\label{lc_pdc_best_fit_no_reg}
\end{figure*}
\begin{figure}[ht]
\begin{center}
\includegraphics[width=0.34\textwidth,trim={2.5cm 2.5cm 2.5cm 3.7cm},clip,angle=90]{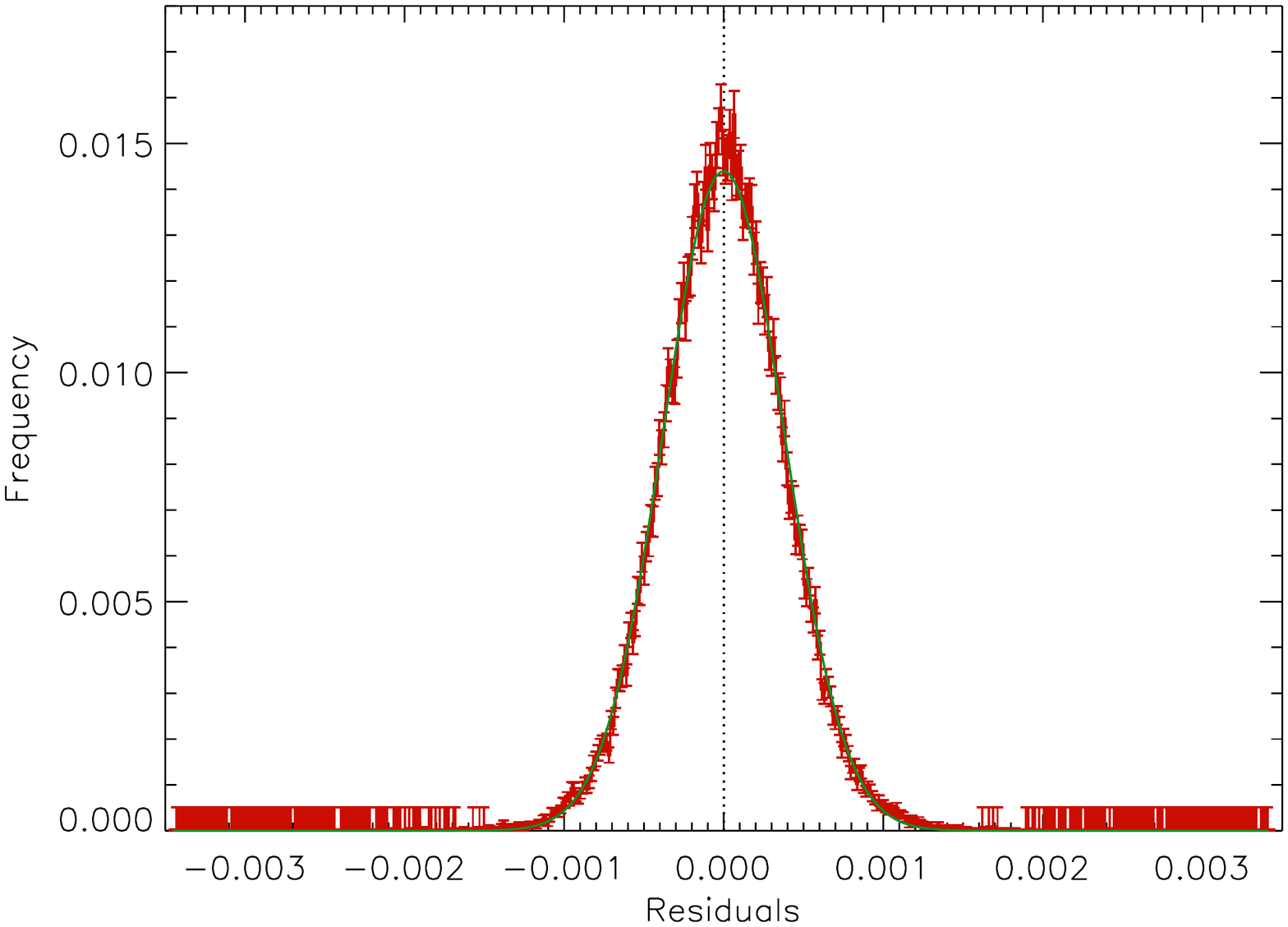}
\end{center}
\caption{Distribution of the residuals of the unregularized best fit to the PDC light curve in Fig.~\ref{lc_pdc_best_fit_no_reg}.}
\label{lc_pdc_best_fit_no_reg_resid}
\end{figure}
\begin{figure}
\begin{center}
\includegraphics[width=0.36\textwidth,trim={2.5cm 2.5cm 2.5cm 4.5cm},clip,angle=90]{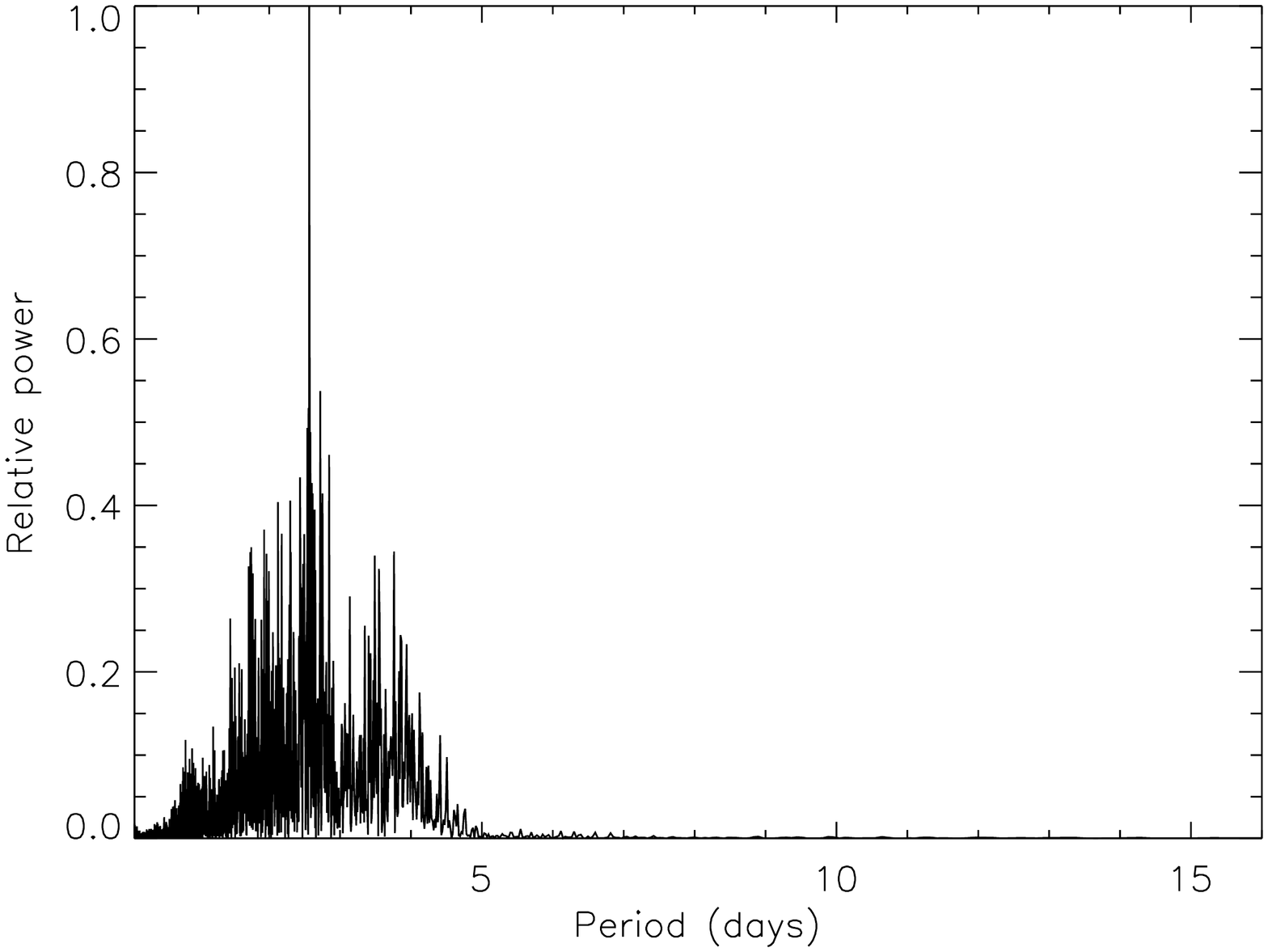}
\end{center}
\caption{GLS periodogram of the residuals of the unregularized best fit to the PDC light curve in Fig.~\ref{lc_pdc_best_fit_no_reg}. }
\label{lc_pdc_best_fit_no_reg_resid_gls}
\end{figure}

In Fig.~\ref{lc_pdc_best_fit_no_reg} we plot the PDC light curve together with the unregularized best fit obtained with our model and the corresponding residuals. The unregularized best fit provides  the minimum chi square and effectively reproduces the spot light modulation giving residuals where it has been filtered out. 

The distribution of the residuals is shown in Fig.~\ref{lc_pdc_best_fit_no_reg_resid} and is almost  Gaussian with a mean $\mu_{\rm res}$ very close to zero ($\mu_{\rm res} = -2.18 \times 10^{-7}$) and a standard deviation $\sigma_{0} = 3.78 \times 10^{-4}$ in relative flux units, slightly larger than the photon shot noise. This indicates the presence of other processes that contribute to the random noise level, probably associated with surface convection and magnetic fields affecting the stellar flux. 

A GLS periodogram of the residuals is shown in Fig.~\ref{lc_pdc_best_fit_no_reg_resid_gls}. Our spot model strongly reduces any variability with periods longer than about 6~days leaving the variability below $3-4$ days almost intact. The maximum of the GLS is reached at a period of $2.565$~days. The peak is about two times higher than the noise level as indicated by the other nearby peaks, thus it is probably not very significant -- usually a peak reaches a significance of 99 percent when its height is at least four times the noise level --, but we shall discuss its possible origin in Sect.~\ref{multi_scale_fluct}.

The unregularized best fit of the SAP light curve is very similar to that of the PDC light curve except for a few larger residuals probably associated with some small uncorrected instrumental effects. By fitting the residual distribution with a Gaussian, we find a slightly larger mean $\mu_{\rm reg} = - 1.737 \times 10^{-6}$, but a standard deviation  $\sigma_{0} = 3.746 \times 10^{-4}$ in relative flux units virtually identical to that of the PDC residual distribution. Also the periodogram of the residuals is very similar and is not shown here.   
The residuals of the PDC and SAP light curve fits will be used to study the multifractal properties of the stellar flux variability without the additional complication introduced by the light modulation produced by starspots. Analysing the residuals of the unregularized best fits will allow us to avoid possible systematic effects introduced by the a priori assumptions adopted for the ME regularization that increase the standard deviation and make the mean of the residuals systematically negative (see Appendix~\ref{appendix_spot_model}). 

\subsection{Spot maps and variation of the spotted area}
\label{spotted_area_var}
\begin{figure*}
\begin{center}
\includegraphics[width=0.68\textwidth,trim={2.5cm 2.5cm 2.5cm 2.5cm}]{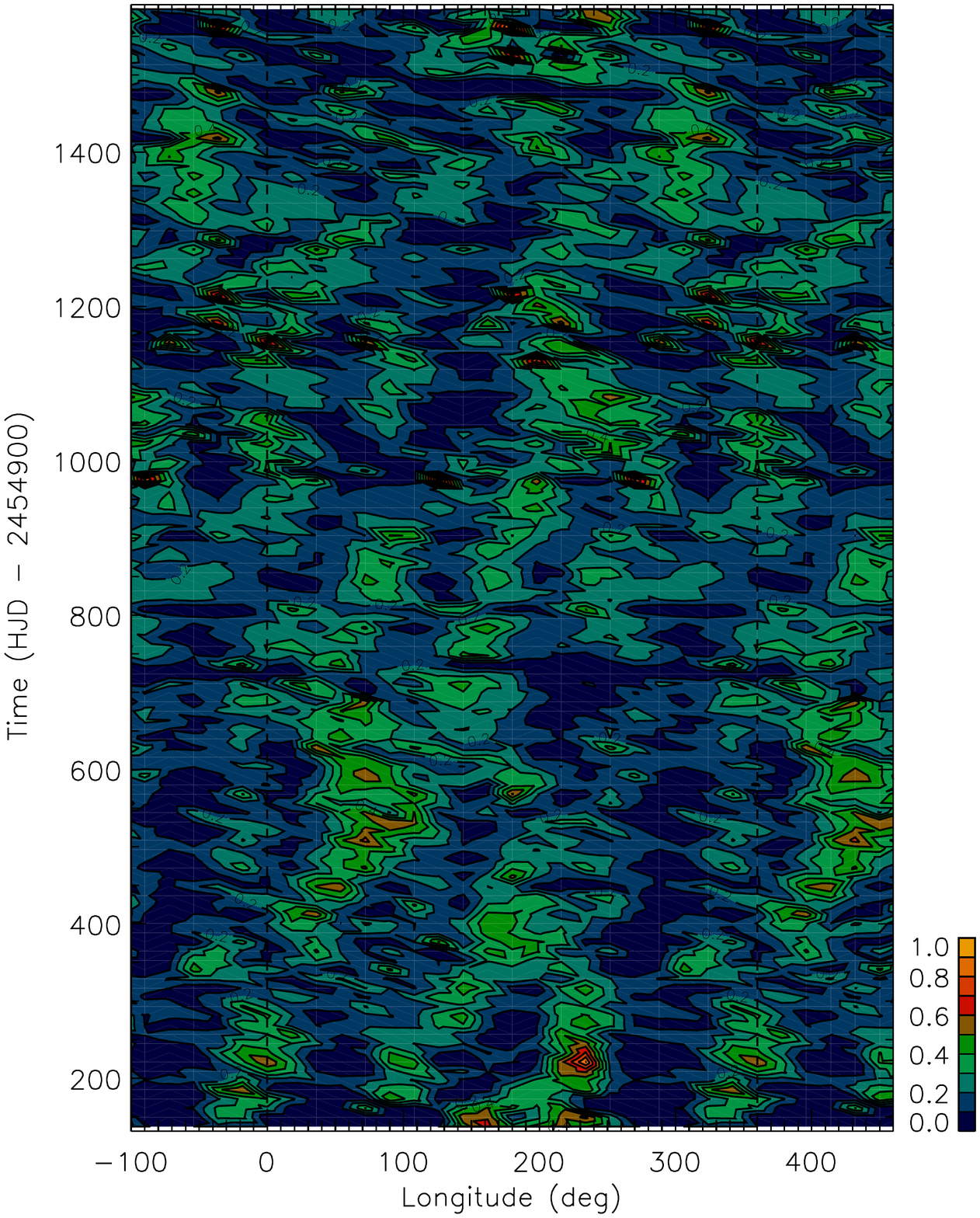}
\end{center}
\caption{Distribution of the spot filling factor vs. the longitude and time for the ME regularized spot model  of the PDC light curve of Kepler-30. The minimum of the filling factor corresponds to dark blue regions, while the maximum is rendered in orange (see the colour scale close to the lower right corner of the plot). Note that the longitude scale of the horizontal axis is extended beyond the $[0^{\circ}, 360^{\circ}]$ interval to better follow the migration of the starspots. }
\label{spot_map_pdc}
\end{figure*}

The spot map obtained from the ME regularized best fit of the PDC light curve is plotted in Fig.~\ref{spot_map_pdc}, where we plot the spot filling factor vs. the longitude and the time. The adopted longitude reference frame rotates with the mean rotation period of the star of 16.0 days and the longitude increases in the direction of stellar rotation. Therefore, spots rotating with the mean rotation period of 16~days stay at a constant longitude on this plot. On the other hand, spots rotating faster than the mean rotation period show a longitude that increases vs. the time, while those rotating slower show a longitude decreasing in time. 
We see that the evolution of the starspots of Kepler-30 is rather fast with sizeable variations of the filling factor over timescales of 20-30 days (the time resolution of our spot modelling is $\Delta t_{\rm f} = 11.963$~days). We introduce a modified Julian date  as MJD $=$ HJD $-2454900$. During the first half of the modelled time series, i.e., up to MJD $\sim 800$, spots cluster around two main active longitudes at $\sim 0^{\circ}$ and $\sim 200^{\circ}$, while a smaller and intermittent active longitude is present in between them around $\sim 100^{\circ}$. Starting from MJD~$\sim 300$, the  main active longitudes migrate towards greater and lower longitudes, respectively, probably as a consequence of the decay of the spots rotating with the mean rotation period and the formation of new spots that rotate faster and slower, respectively. The intermediate active longitude is approached by the main longitude initially at $\sim 200^{\circ}$ and looses its identity. For MJD$~\ga 800$, the active longitudes are less well defined and the pattern is characterized by individual spots with lifetimes ranging from $\approx 30$ to $\approx 200$ days that rotate at different rates. The spot map obtained with the regularized best fit to the SAP light curve is similar to that displayed in Fig.~\ref{spot_map_pdc} and is plotted in Appendix~\ref{appendix_spot_model}. 

The migration of the different spots in longitude can be used to estimate the surface differential rotation, if we assume that the migration is caused by the location of those spots at different latitudes on a differentially rotating star. Given the lack of information on the spot latitudes, we can only derive a lower limit to the surface shear. The estimate should be regarded with great caution because the apparent migration can be the result of the starspot evolution rather than of the differential rotation. In the case of Kepler-30, the evolution of individual starspots occurs on timescales shorter than or comparable with those of the migration of the active longitudes, thus this source of systematic errors is certainly present and casts a further uncertainty on our estimate. 

Looking at the spot map in Fig.~\ref{spot_map_pdc}, the best suitable feature to estimate the differential rotation is the active longitude that starts its migration at MJD $\sim 400$ at a longitude of $\sim 30^{\circ}$ and reaches a longitude of $\approx 100^{\circ}$ at MJD $\sim 550$. Taking into account an uncertainty of at least $\pm\, 40^{\circ}$ in the longitude shift between those two dates owing to the intrinsic width and spreading of the active longitude in time, we estimate a relative surface shear of $\Delta \Omega / \Omega \sim 0.020 \pm 0.012$, where $\Omega$ is the angular velocity of rotation, the mean value of which corresponds to the mean rotation period of 16 days. A similar estimate based on the spot map obtained from the SAP timeseries (cf. Fig.~\ref{spot_map_sap}) gives a comparable result. In any case, we stress that this estimate is particularly uncertain, especially because of the fast spot evolution seen inside the active region with individual spot lifetimes of tens of days, i. e., comparable with the time resolution of our spot modelling.  Other active longitudes show a less clearly defined migration pattern and are more affected  by the intrinsic evolution of the spots; thus, they are not considered for our estimate. 

\begin{figure}
\begin{center}
\includegraphics[width=0.33\textwidth,trim={2.5cm 2.5cm 2.5cm 3.5cm},clip,angle=90]{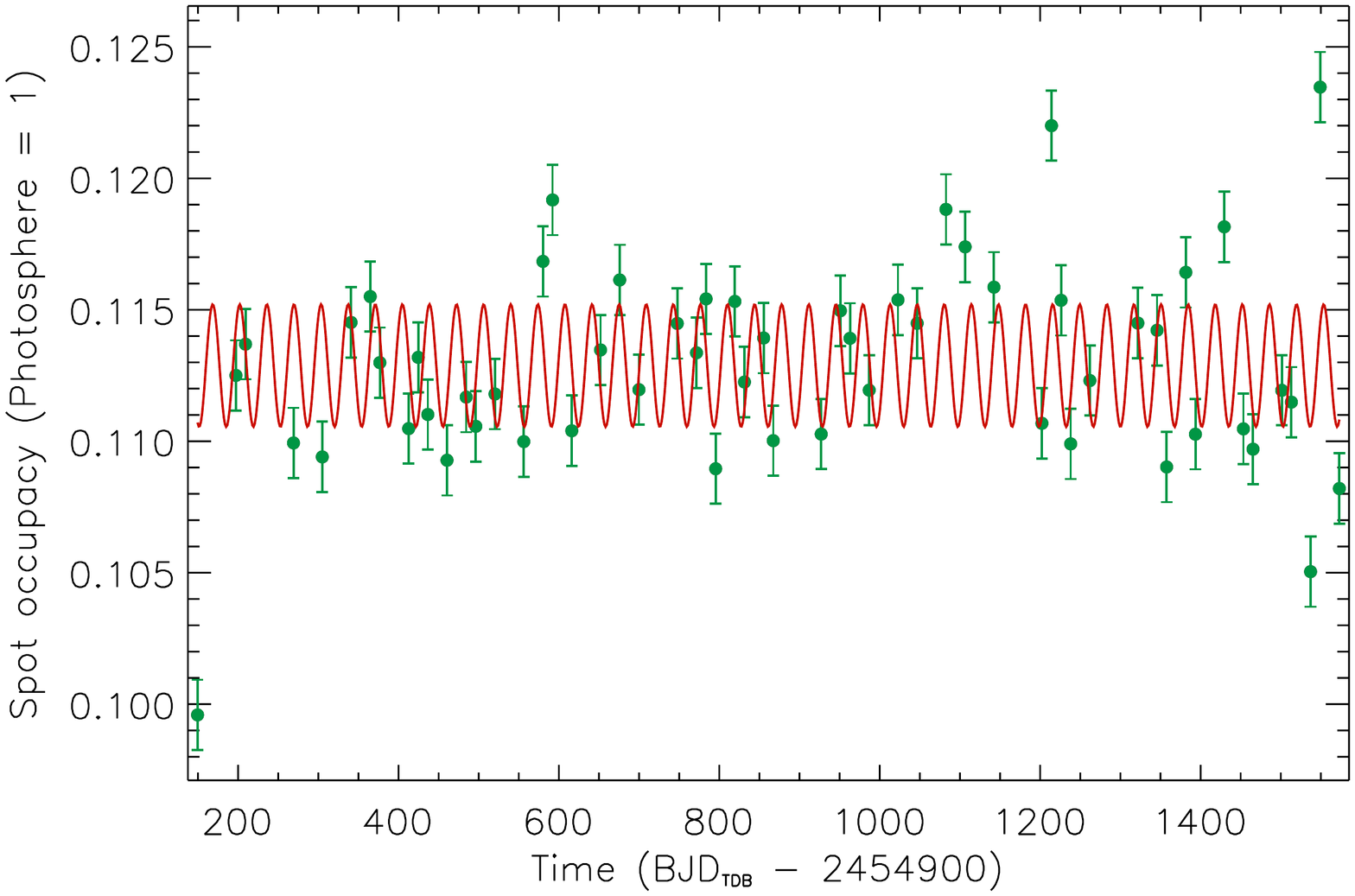} 
\end{center}
\caption{The total spotted area on Kepler-30 as derived from the spot modelling of the PDC light curve vs. the time (green dots). A sinusoid with a period of 33.784~days corresponding to the maximum of the GLS periodogram is overplotted (red solid line). }
\label{total_spot_area}
\end{figure}

\begin{figure*}
\begin{center}
\includegraphics[width=0.68\textwidth]{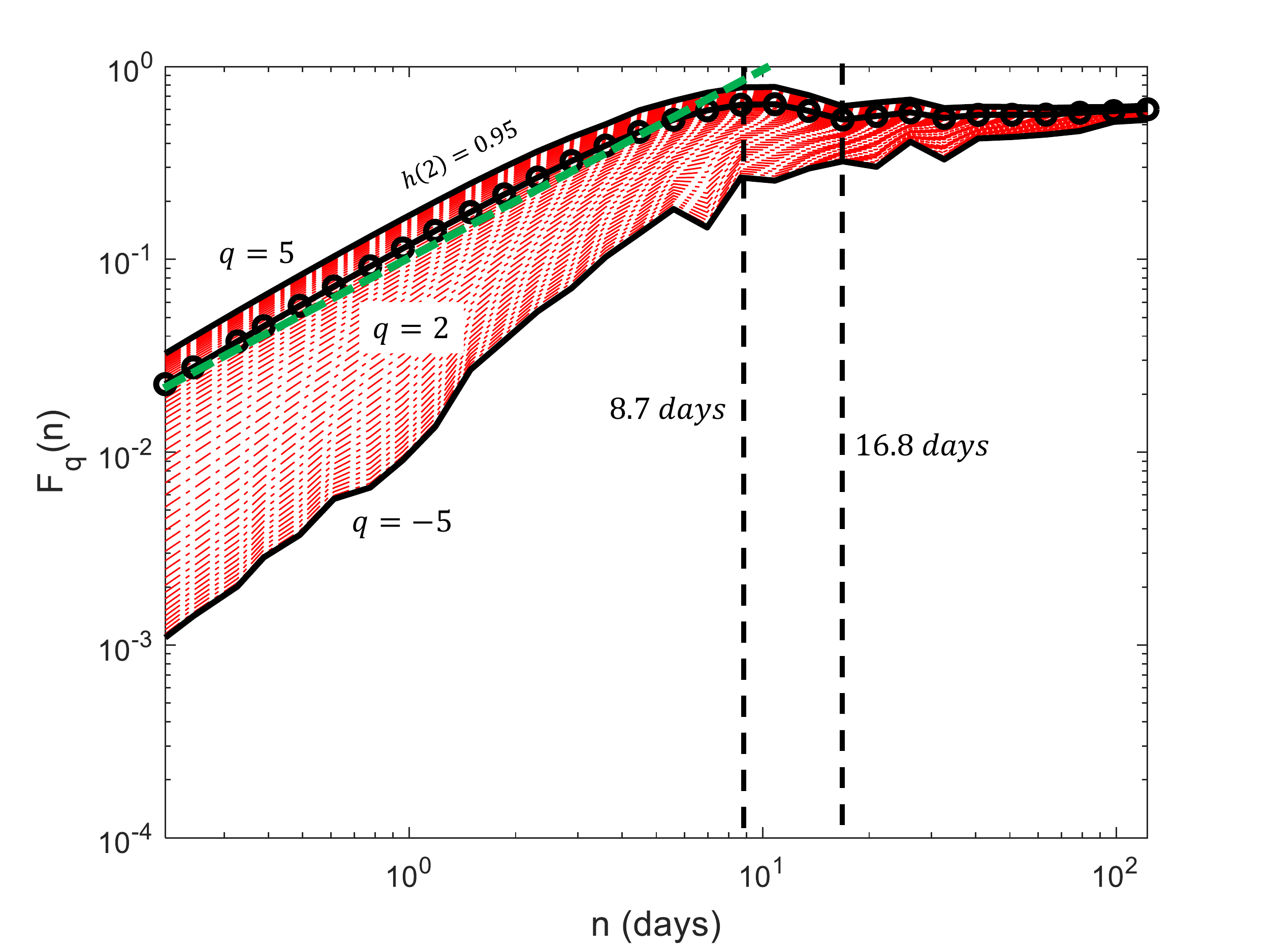}
\end{center}
\caption{The log-log plot of the fluctuation functions $F_{q}(n)$ (black circle for $q=2$) calculated for the final PDC time series presented in Fig. \ref{figLS}. The red curves correspond to $q$ between $-$5 and 5 in steps of 0.2. Vertical dashed lines mark three domains of the fitting windows for the small $n$ between 29.4 min and 8.7 days, for middle $n$ between 8.7 and 16.8 days and, for the large $n$ greater than 16.8 days. The green dashed line gives the average slope $H=h(2)=0.95$ for timescales shorter than 8.7 days.}
\label{figFlu1}
\end{figure*}

\begin{figure*}
\begin{center}
\includegraphics[width=0.68\textwidth]{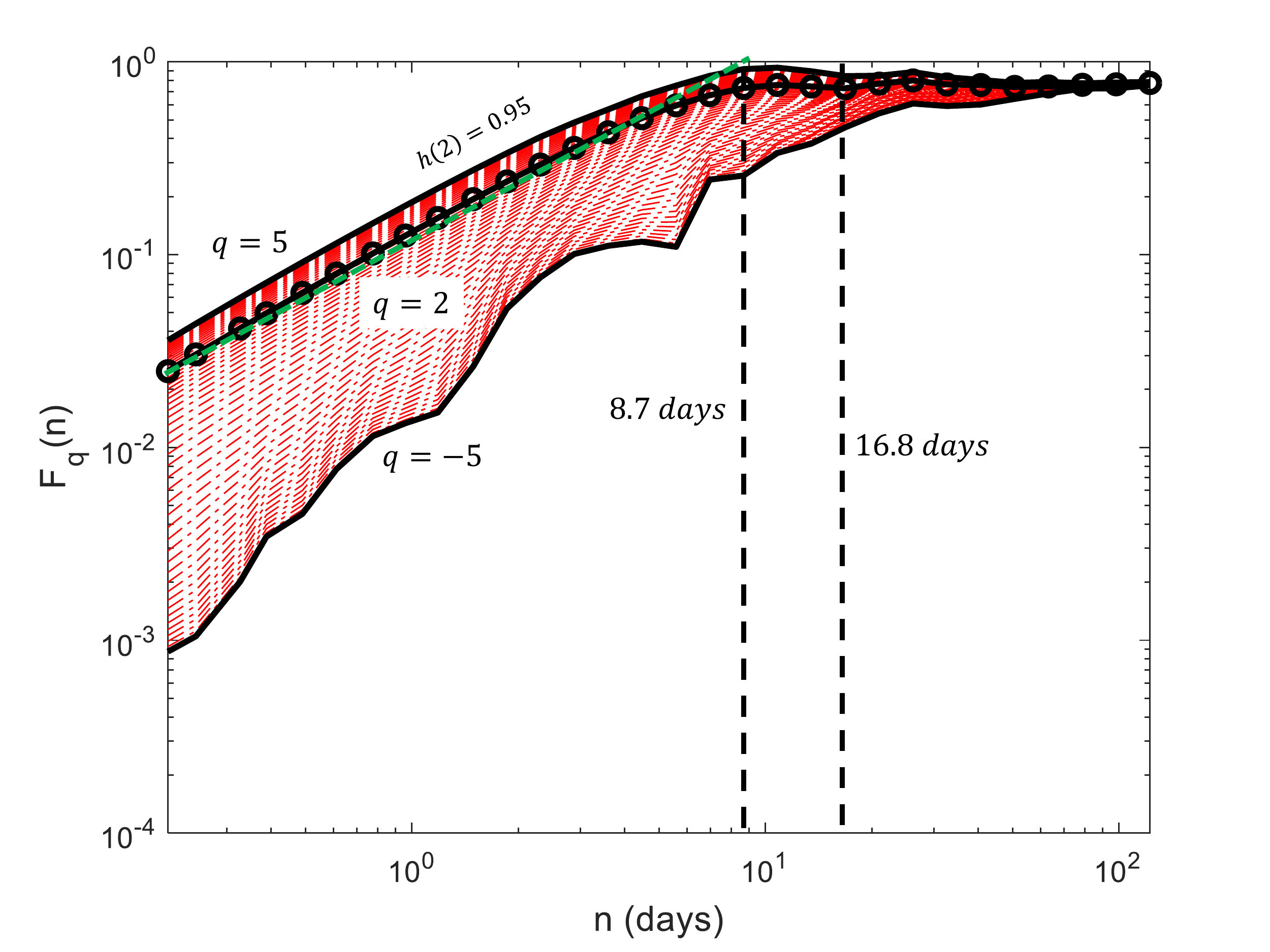}
\end{center}
\caption{Idem Figure \ref{figFlu1} for final SAP time series.}
\label{figFlu2}
\end{figure*}

\begin{figure*}
\begin{center}
\includegraphics[width=0.68\textwidth]{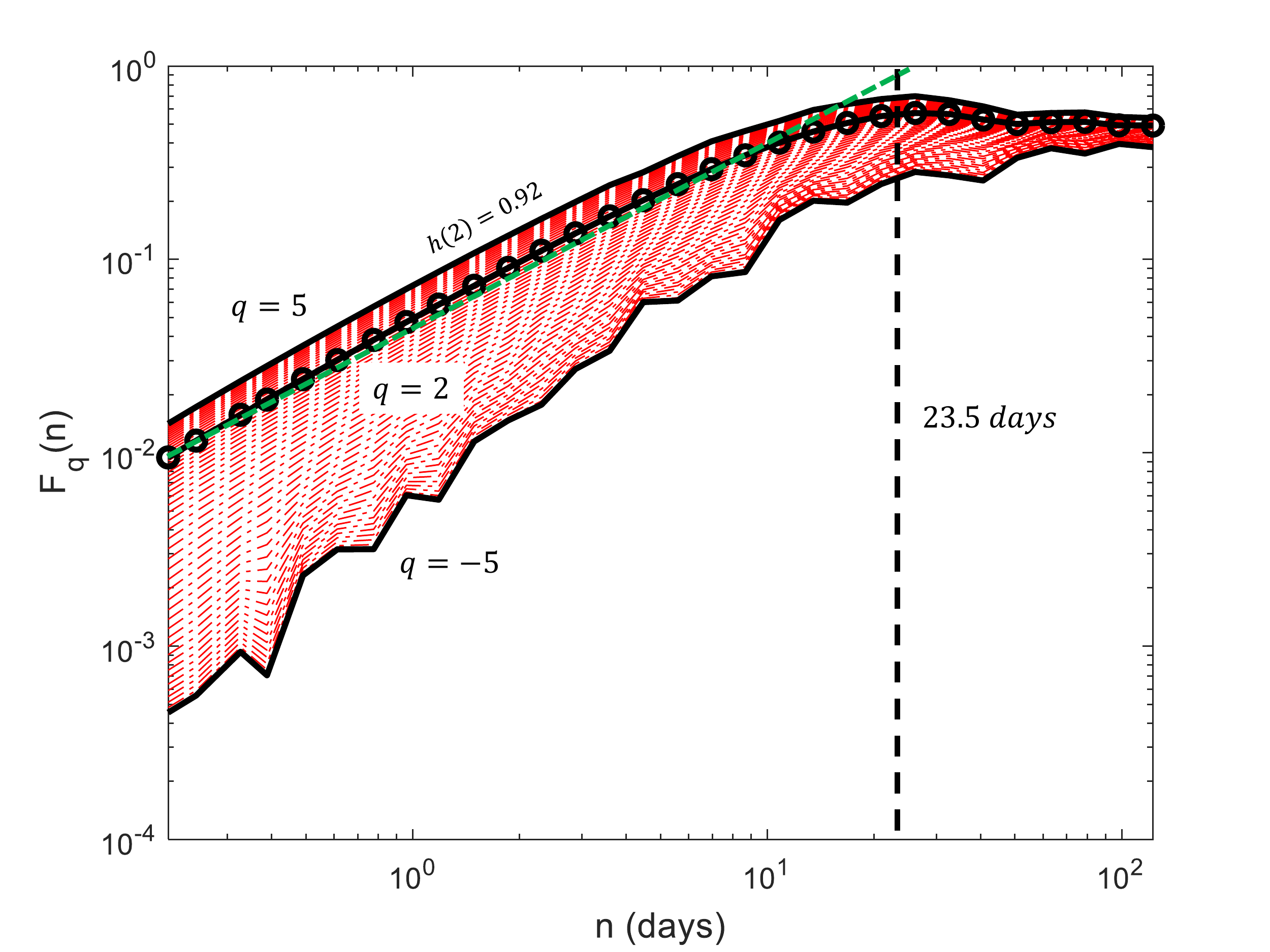}
\end{center}
\caption{Idem Figure \ref{figFlu1} for final RTS time series. In this case, only two domain are considered and separated by the vertical dashed line at 23.5 days.}
\label{figFluResidual}
\end{figure*}


The total spotted area, obtained by summing up the filling factor over all the longitudes, is plotted in Fig.~\ref{total_spot_area} vs. the time. Only the values obtained from time intervals $\Delta t_{\rm f}$ without significant data gaps have been plotted in Fig.~\ref{total_spot_area} to avoid systematic errors associated with the ME regularization that tends to decrease the spotted area in the time intervals containing gaps. To find if the distribution of the datapoints within each interval $\Delta t_{\rm f}$ is uniform, we subdivide it into five equal subintervals  and count the number $q_{i}$ of datapoints in each subinterval ($i =1, ..,5$). If the ratio $(\max\{q_{i}\} - \min\{ q_{i} \})/\max \{q_{i} \} \le 0.2$, the corresponding interval $\Delta t_{\rm f}$ is considered as not affected by gaps and its area value is considered in our analysis.  In spite of disregarding the area values obtained from intervals with significant gaps, we still see some points with deviations up to $\pm 10$ percent from the mean value in Fig.~\ref{total_spot_area}. Such relatively large deviations are due to flux discontinuities at the epochs when successive intervals are stitched together after a repointing of the Kepler telescope or to some small deviations from the convergence criterion adopted to fix the level of regularization (see Appendix~\ref{appendix_spot_model}).

A GLS periodogram of the spotted area shows a peak at 33.8~days with a false-alarm probability (FAP) of 0.29 according to the analytical formula by \citet{ZechmeisterKuerster09}. The corresponding best fitting sinusoid is plotted in Fig.~\ref{total_spot_area} with the area values affected by the systematic effects mentioned above clearly deviating from it. Repeating the same analysis with the spot model of the SAP light curve (see Appendix~\ref{appendix_spot_model}), not affected by the systematic effects introduced by the PDC pipeline, we find the same periodicity with a period of 33.9~days and a FAP $= 0.024$, thus confirming the presence of a short-term modulation in the total spotted area of Kepler-30, reminiscent of the short-term cycles found in CoRoT-2 \citep{Lanzaetal09,Zaqarashvilietal21}. 

Such a short-term cycle of $\sim$34 days is virtually close to the synodic period $P_{\rm syn}=35.2$~days of the planet Kepler-30b as computed with a rotation period $P_{\rm rot} = 16.0$ days and an orbital period $P_{\rm orb} = 29.334$~days from $P_{\rm syn}^{-1} = P_{\rm rot}^{-1} - P_{\rm orb}^{-1}$. This period also appears in the periodogram of Fig.~\ref{figLS2} and is close  to the period of the modulation of the spotted area.  In addition, the 23.1-day period (see Fig.~\ref{figLS2}) is close to the synodic period of Kepler-30c. However, an interpretation in terms of tidal or magnetic star-planet interactions \citep[cf.][]{Lanza12,Lanza13} is problematic because of the large star-planet separations that range from $\sim 42$ to $\sim 121$ stellar radii for the three planets (see http://exoplanets.org). More precisely, both the tidal torque on the star and the energy density of its coronal magnetic field decrease with distance as $(a/R_{\rm s})^{-6}$, where $a$ is the orbital separation and $R_{\rm s}$ the radius of the star.  

\subsection{Multiscale multifractal analysis}
\label{multi_scale_analysis}
\subsubsection{Fluctuation functions}
\label{multi_scale_fluct}

In Figs.~\ref{figFlu1} and~\ref{figFlu2}, we plot the fluctuation functions $F_{q}(n)$ vs. the timescale $n$ for different values of $q$ for the PDC and SAP timeseries, respectively. In this logarithmic plot, the scaling relation given by equation~(\ref{Fqxn}) becomes a straight line in the intervals of $n$ where $h(q)$ is constant. A value of $n$ separating two consecutive intervals where $h(q)$ is constant for a given $q$ is called a {\em crossover}. We see a crossover in the plots for the values of $q>0$ at $8.7$ days as indicated by the vertical dashed line. The crossover corresponds to the optimal delay time as determined with the procedure described in Section~\ref{tau} (cf. Fig.~\ref{figFirst3}). The change in slope is made clear by the green dashed line that gives an average slope $H=h(2)=0.95$ for timescales shorter than 8.7 days. The small decrease in the fluctuation functions following the crossover and extending up to the rotation period of $16.8$ days is barely significant and is due to the variation in amplitude produced by the use of a moving average to approximate what is actually a nearly sinusoidal modulation in the computation of the fluctuation function itself (see Appendix~\ref{multifractal_background}). Once the rotation period is reached, the amplitude of the fluctuation function saturates at a nearly constant value, except for some small oscillations produced by the sinusoidal modulation of the signal (cf. Appendix~\ref{fluc_func_sinusoid}). In other words, the slope $h$ beyond the crossover is virtually zero because the variability due to the rotational modulation dominates over the stochastic fluctuations. We remind that the fluctuation function with $q=2$ is based on the variance of the fluctuations in the time series.  

For both the PDC and SAP timeseries, we find a crossover at the timescale corresponding to  the first harmonic of the rotation period, i.e., $\sim 8.7$~days. The predominance of the modulation at the first harmonic is seen in the light curves of active stars when two active longitudes on opposite hemispheres are responsible for most of the flux modulation, a behaviour that is confirmed by our spot modelling (see Sect.~\ref{spotted_area_var}). From  the perspective of stochastic process analysis, the behaviour we see is reminiscent of an attractor (see Figure~\ref{figFirst3}) with a phase trajectory that revolves around a period-one fixed point. As a result, the logarithms of the fluctuation functions $F_{q}(n)$ show a linear regime until $\sim$8.7 days after which they get saturated and start weakly oscillating. 
There is a slight difference in the level of oscillation when the time series PDC and SAP are compared as can be seen in Figures~\ref{figFlu1} and \ref{figFlu2}. Based on the results in Sect.~\ref{best_fit}, the presence of a few larger residuals found in the SAP time series probably explains the plateau within the range from 8.7 to 16.8 days, since the noise is responsible for dampening the oscillation.

In Fig.~\ref{figFluResidual}, we plot the fluctuation functions for the difference time series RTS. In this case, we see a crossover at a longer timescale of $\sim 23.5$ days. We can associate it  with the characteristic evolutionary timescales of active regions in Kepler-30 because the rotational modulation was remarkably reduced by subtracting the two timeseries from each other (cf. Sect.~\ref{spotted_area_var}). We remind that the intrinsic variability on timescales longer than $15-20$ days was preserved in the SAP timeseries, while it was strongly reduced in the PDC timeseries owing to the de-trending applied by the Kepler pipeline (cf. Sect.~\ref{data_preparation}). Therefore, the long-term intrinsic variability stands out in the difference timeseries. 

\begin{figure*} 
\begin{center}
\includegraphics[width=0.68\textwidth]{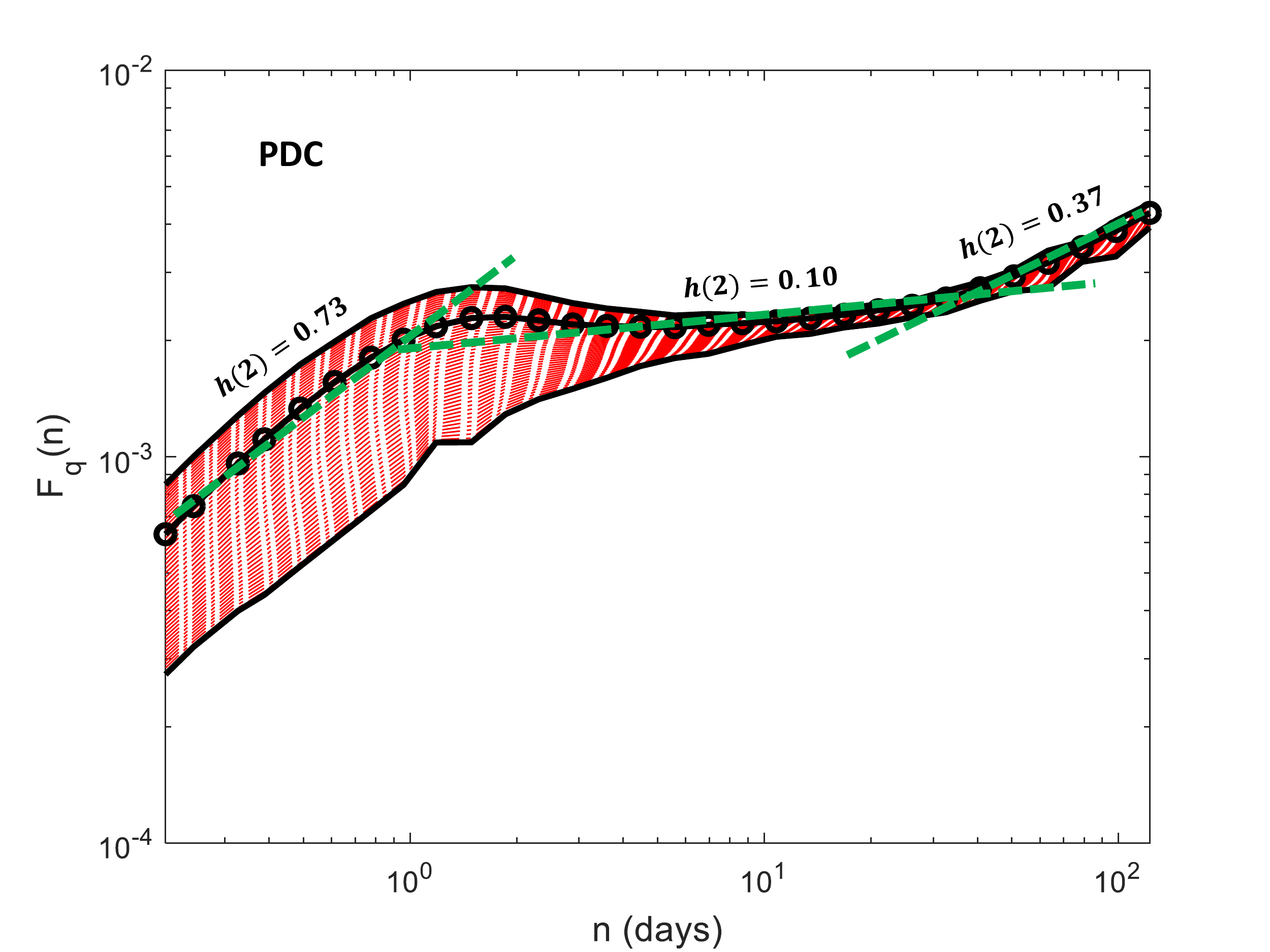}
\end{center}
\caption{Fluctuation functions of the residuals of the unregularized spot model of the PDC light curve of Kepler-30.}
\label{multi_fractal_PDC_spot_res}
\end{figure*}

The fluctuation functions of the timeseries of the residuals of the unregularized spot modelling of the PDC light curve are plotted in Fig.~\ref{multi_fractal_PDC_spot_res}.  We see two crossover at $\sim 1$ and $\sim 35$~days, where the Hurst exponent $H = h(2)$ changes from $0.73$ to $0.10$ and then to $0.37$, respectively. Therefore, the fluctuations with timescales shorter than $\sim 1$~day are characterized by persistence, while those on longer timescales show a value of $H$ characteristic of antipersistent time series. The timescale of $\sim 1$~day is characterized by a remarkable increase of the power level in the GLS power spectrum of the residuals (cf. Fig.~\ref{lc_pdc_best_fit_no_reg_resid_gls}) and the multifractal analysis reveals a change in the persistence property of the fluctuations just at that timescale. It corresponds to the characteristic turnover time of supergranular convective cells in the Sun and Sun-like stars \citep[cf.][]{Meunieretal15}. The other crossover at $\sim 35$ days coincides with the possible short-term cycle in the area of the starspots (cf. Sect.~\ref{spotted_area_var}) and the change in the slope of the fluctuation functions may indicate changes in the properties of convection at that timescale of stellar activity.  The fluctuations functions of the spot model residuals of the SAP lightcurve are very similar and shall not be discussed here.  In both the cases, we do not find any particular feature of the fluctuation functions at the peak period of 2.565~days in the GLS periodogram of the residuals, suggesting that it is not significant and not associated with a change in the behaviour of the stellar light fluctuations. 

\begin{figure*}
\begin{center}
\includegraphics[width=0.68\textwidth]{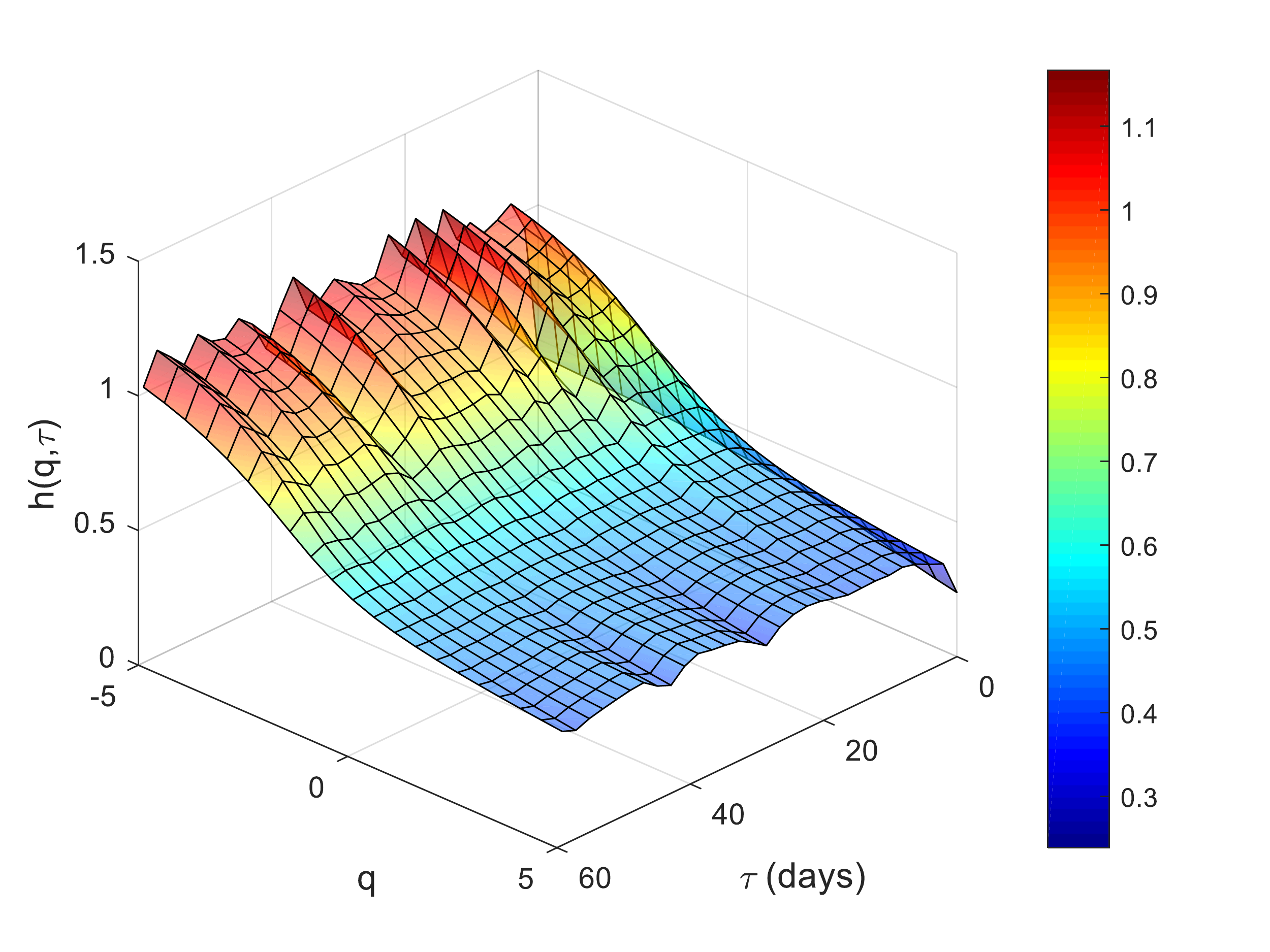}
\end{center}
\caption{Hurst surface $h(q,\tau)$ calculated for final PDC time series. The $(h,q)-$plane corresponds to $h(q)$ calculated with the standard MFDMA method. Colorbar indicates values of  $h(q,\tau)$, where for $q<0$ the higher values are found and, for $q>0$ the lower ones.}
\label{figHqt1}
\end{figure*}

\begin{figure*}
\begin{center}
\includegraphics[width=0.68\textwidth]{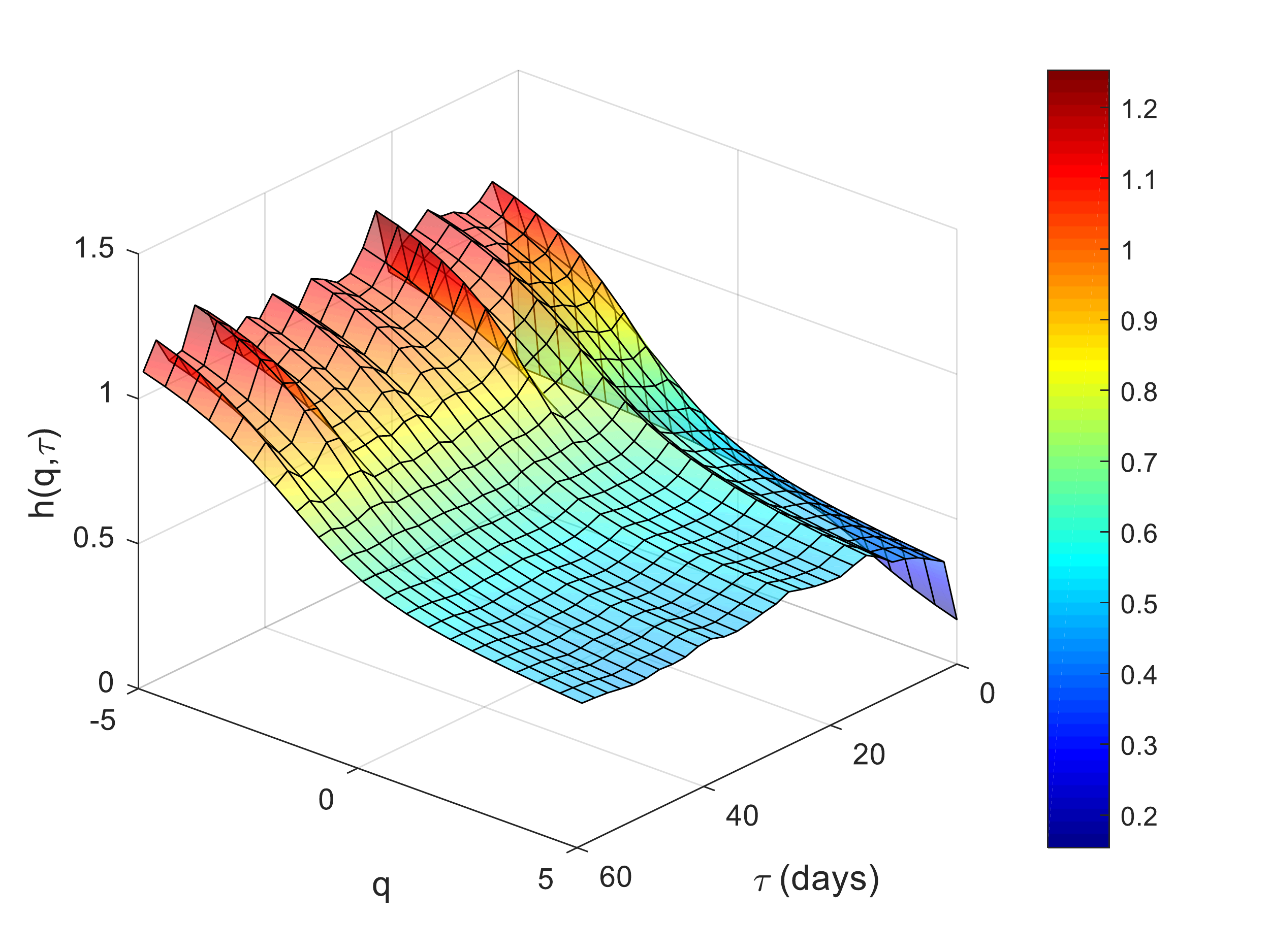}
\end{center}
\caption{Idem Figure \ref{figHqt1} for SAP data.}
\label{figHqt2}
\end{figure*}

\begin{figure*}
\begin{center}
\includegraphics[width=0.68\textwidth]{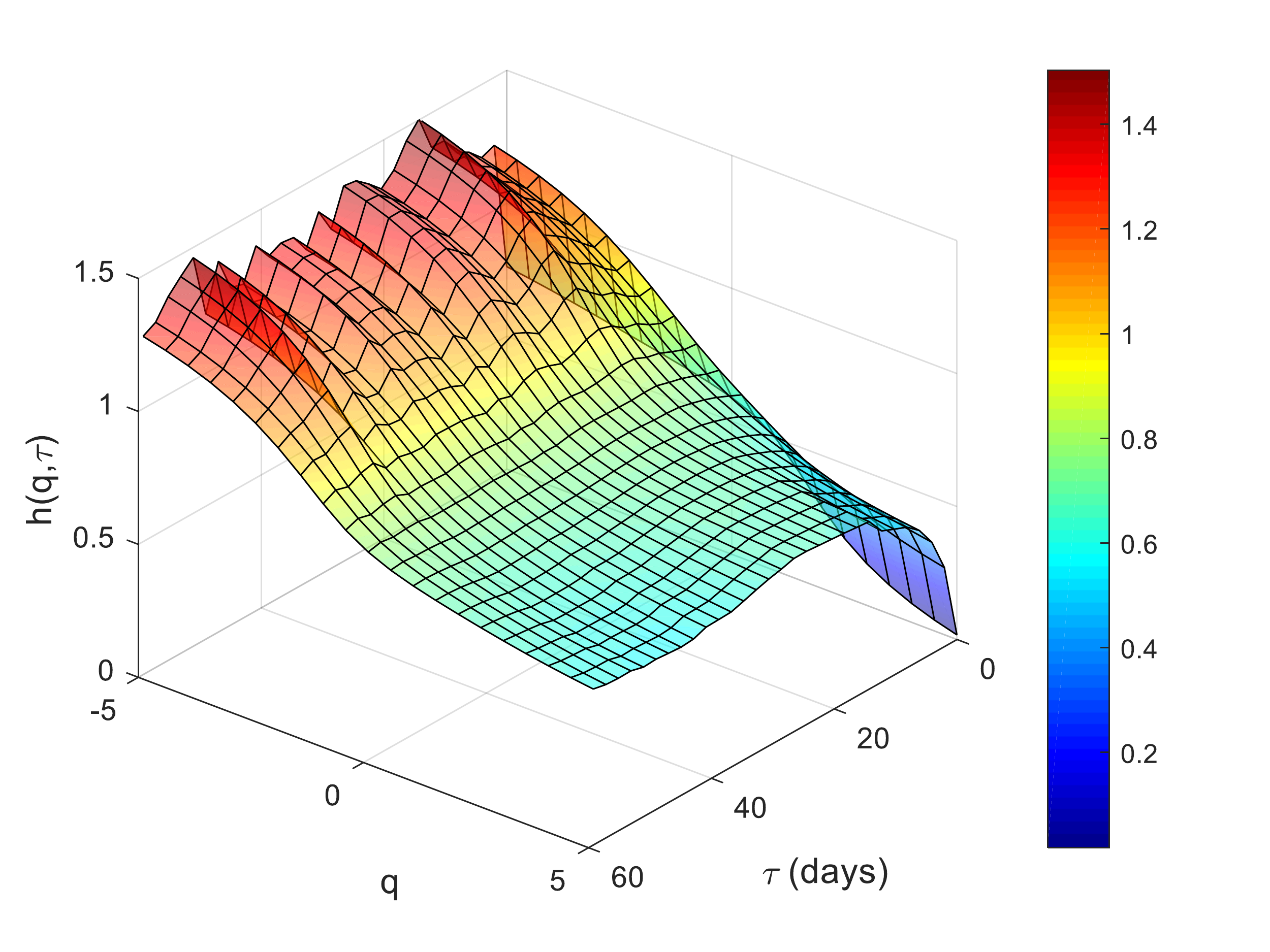}
\end{center}
\caption{Idem Figure \ref{figHqt1} for RTS data. In this case, random walk-type fluctuations ($H>1$) are stronger than PDC and SAP data, indicating more apparent slow evolving fluctuations in regime $q<0$.}
\label{figResidual}
\end{figure*}

\begin{figure*}
\begin{center}
\includegraphics[width=0.8\textwidth]{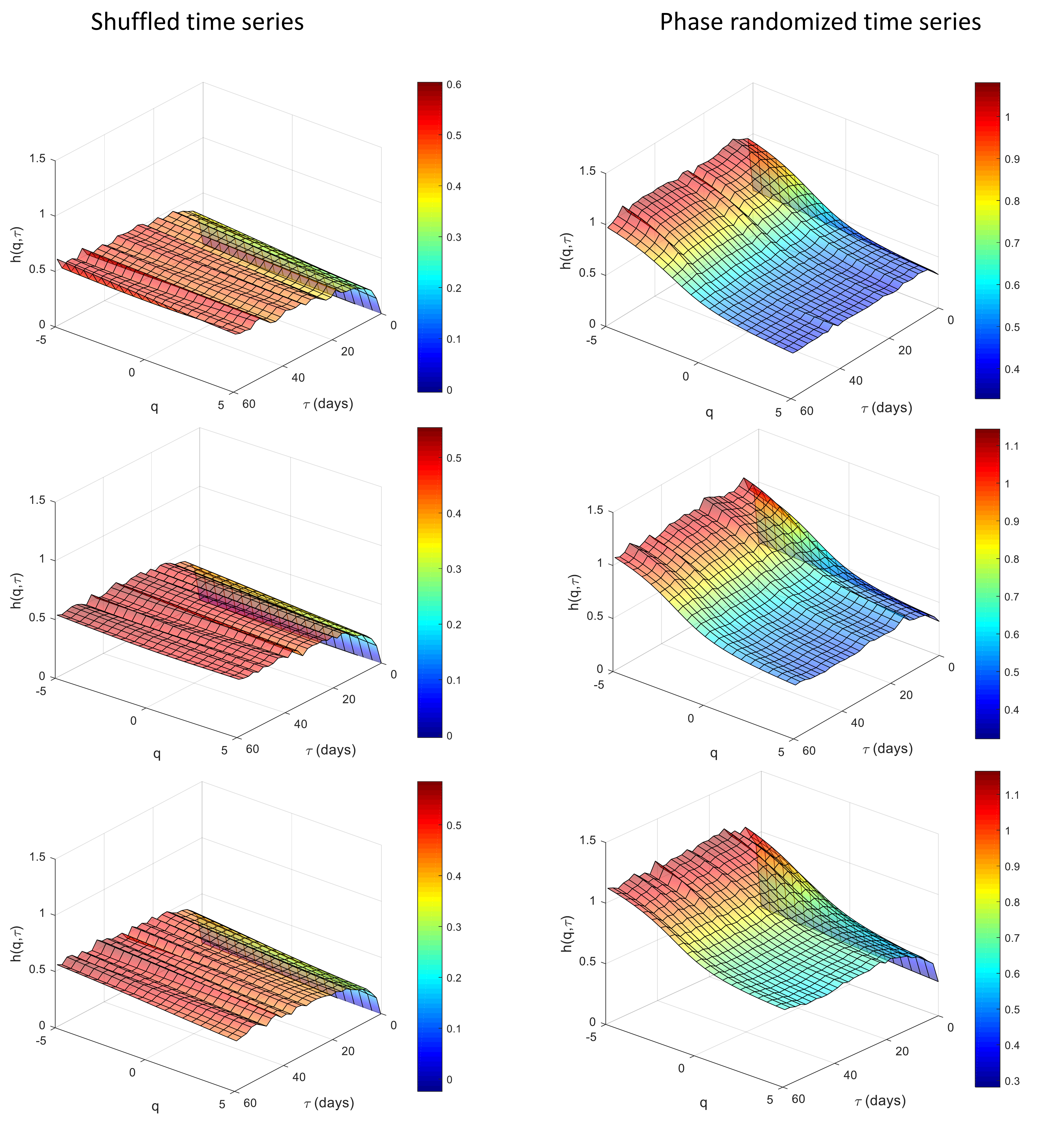}
\end{center}
\caption{Hurst surface $h(q,\tau)$ calculated for the shuffled (left side) and phase randomized (right side) versions of time series: PDC (top), SAP (middle) and RTS (bottom). Presented are results averaged over 200 realizations of the surrogate series.}
\label{figSP}
\end{figure*}

We based our conclusions on the fluctuation functions $F_{q}(n)$ with positive $q$ where the effect of large-scale fluctuations is amplified with respect to that of small-scale fluctuations in the computation of the moments of order $q>0$. Conversely, small-scale fluctuations will have the most important impact on the fluctuation functions $F_{q}(n)$ when $q< 0$ because the effect of large-scale fluctuations is attenuated by a negative exponent in the calculation of the momenta (cf. eq.~\ref{eq4}). In all the above plots of the fluctuation functions $F_{q}(n)$ with $q <0$, we see that the exponent of the scaling law, that is, the slope of the plot, changes frequently and abruptly. This indicates that the asymptotic regime where $F_{q} (n) \propto n^{h(q)}$ is not fully reached in those plots likely as an effect of a few very small fluctuations that dominate the functions. 

\subsubsection{Hurst exponent surfaces}

Now, we pass to investigate the effects of a periodic trend using the Hurst surface $h(q, \tau)$. With this, we can  study the generalized Hurst exponent $h$ behaviour not only as a function of the $q$-moment (standard MFDMA method) but also as a function of the timescale $\tau$ (MFDMA$\tau)$. The Hurst surfaces illustrated in Figs. \ref{figHqt1}, \ref{figHqt2} and \ref{figResidual} calculated for PDC, SAP and RTS data show abundant features to analyse, which may be hidden by the standard MFDMA proposed by \cite{gu2010}.

For a fixed scale $n$, when $q$ changes from $-$5 to 5, there are downward trends for all $h(q,\tau)$ surfaces, except in the domain where $n<8.7$ ($n<23.5$) days and $q<0$ for PDC and SAP (RTS) series. In particular, we can recover the results from the standard MFDMA method, calculating the scaling exponent for each $q$ at the whole scale range $n$. In this way, varying $q$, when $\tau$  changes from small  to large values, all curves of $h(q>0,\tau)$ show a rapid rise to their peak values, then become more or less constant showing a plateau.  However, for $h(q<0,\tau)$, there are oscillations that increase with the increase of $|q|$, in addition to a stronger jump when $\tau$ is small and $q<0$ for PDC and SAP and less pronounced for RTS. The presence of bumps at negative $q$ values in the Hurst surfaces in Figs. \ref{figHqt1}, \ref{figHqt2} and \ref{figResidual} may be the results of the amplified impact of small fluctuations that in turn depends on several effects such as the evolution of active regions and the differential rotation. 

The main feature of the $h(t, \tau)$ plots is the remarkable decrease of the value of $h$ from $\approx 1$ for $q \sim -5$ to $\sim 0.5$ for $q >0$. This suggests that the small scale fluctuations, the effect of which on $h$ is amplified for $q<0$, have a strong degree of persistence, while the large scale fluctuations, which dominate $h$ for $q>0$ are more similar to an uncorrelated random process. The results based on the portion of the plots with $q < 0$ should be taken with some caution because of the sizeable fluctuations of $h$, but these plots suggest a different physical origin for small and large scale fluctuations in the Kepler-30 light curve, a conclusion that can be useful to constrain models of stellar microvariability. Crossovers are not clearly evident in the $h(t, \tau)$ plots of the SAP and PDC timeseries, except for that at $\tau = 8.7$ days already discussed above. This happens probably because the variability is dominated by the rotational modulation. Therefore the SAP and PDC $h(t,\tau)$ surfaces show only a tiny hint of possible slope changes for $\tau$ between $\approx 20$ and $\approx 35$~days that could be associated with the evolution of active regions. On the other hand, the $h(t,\tau)$ plot of the RTS timeseries shows a clear crossover at $\sim 23$ days that extends over an interval of $q$ and that is likely indicating the typical active regions evolution timescale as estimated from the spot modelling.   

Following the discussion in Sect.~\ref{origins_of_mf} and in \cite{defreitas2017}, we computed the $\langle S_{\rm ph} \rangle_{k=5}$ activity index for Kepler-30 finding 4712  and 5209 parts per million for the PDC and SAP pipelines, respectively. The larger value found with the SAP data is a consequence of the reduction of the stellar variability in the PDC timeseries for timescales longer than $\sim 15$~days. The Hurst exponent surface $h(2,\tau)$ shows a steep increase up to $\tau=8.7$~days followed by an almost constant plateau, indicating that most of the variability in the timeseries of Kepler-30 occurs for timescales shorter than half the rotation period of the star which corresponds to only  10\% of the time interval sampled by the $\langle S_{\rm ph} \rangle_{k=5}$ index. Therefore, such an index provides a very limited description of the  light variability in the case of Kepler-30, while the Hurst  exponent surfaces allow a detailed description of the dependence of the  variability on the timescale. The short-term spot cycle of $\sim 34$~days of Kepler-30 has a period shorter that the five rotations adopted to compute the $\langle S_{\rm ph} \rangle_{k=5}$ index, therefore this index may not be appropriate to sample the activity timescales characteristic of Kepler-30.

\subsubsection{Effect on surrogate time series}
As mentioned in Sect.~\ref{origins_of_mf},  multifractality can have two sources: long-term correlations or a fat-tailed probability distribution of the fluctuations. In the case of Kepler time series, it was already shown that the first source of multifractality is by far the dominant one \citep{defreitas2017}. 

We confirm this result in the specific case of Kepler-30 by calculating $h(q,\tau)$ for the shuffled and phase-randomized surrogates of the three timeseries considered above. 

Figure \ref{figSP} shows the average results of 200 realizations of the shuffled and phase-randomized surrogates of PDC (upper panels), SAP (middle panels) and RTS (bottom panels). In fact, as can be seen in Fig. \ref{figSP}, the shuffling procedure destroyed the correlations, i.e., the Hurst surface $h(q,\tau)$ is flat ($\langle h(q,\tau)\approx 0.5\rangle$, the value of $h$ corresponding to white noise). However, the Hurst surface of phase-randomized series vary slightly both with the order $q$ and timescale $\tau$ ($\langle h(q,\tau)\approx 0.82\rangle$). These findings suggest that the multifractality of rotational modulation is due to both long-range correlation and non-linearity {due to fat-tailed probability distributions}, but  long-range correlations are the main source of multifractality, which is consistent with the results of standard MFDMA. 

\section{Summary and Conclusions}
In this study, we analysed the rotation and the evolution of photospheric active regions of the moderately young Sun-like star Kepler-30 accompanied by a three-planet system, using  both its PDC and SAP timeseries by means of an MFDMA-based multifractality analysis approach in a standard and a new multiscale version. In the latter case, the PDC and SAP timeseries, as well as the difference RTS data, are examined considering the generalized dependence of the local Hurst exponent on the timescale $\tau$ by means of the surface $h(q,\tau)$. We also consider the impact of the rotational modulation on the characterization of the multifractal properties of the light fluctuations, as already investigated by \cite{defreitas2013b,defreitas2016,defreitas2017,defreitas2019a,defreitas2019b}, and show that such an impact depends on the timescale. Furthermore, we applied a maximum entropy spot modelling to extract information on the longitude, the area variation, and the evolutionary timescales of the active regions responsible for the rotational modulation of the stellar flux. Such an approach reveals that the characteristic timescales of stellar activity in Kepler-30 are significantly shorter than the five rotations adopted to compute the $\langle S_{\rm ph}\rangle_{k=5}$ index as defined by \cite{mathur2014a}, therefore such an index is of limited use to characterize the activity of our target.

Our main conclusions can be summarized as follows:

(i) The fluctuation functions $F_{q}(n)$ show that the multifractal properties of the Kepler-30 timeseries have a relationship with the range of scale $n$ and, therefore, indicate the limitation of the standard MFDMA method using a fixed timescale. Then, we systematically investigate the dynamic behaviours of the three timeseries, PDC, SAP and RTS, by applying a new approach here named MFDMA$\tau$;

(ii) The Hurst surfaces reveal that for negative $q$ values there are remarkable fluctuations in the local Hurst exponent $h(q, \tau)$ and significant differences for different values of $q$. Conversely, the positive $q$ values show that the Hurst surfaces become flat starting from a minimum period ($\sim 8.7$~days) that corresponds to the first harmonic of the rotation period as found by the Lomb--Scargle periodogram in the case of the SAP and PDC timeseries, while it corresponds to the evolutionary timescale of active regions ($\sim 23.5$~days) for the RTS timeseries, in which the rotational modulation has been almost completely removed. The analysis of the residuals of the spot modelling shows a crossover at $\sim 1$ day, that coincides with the characteristic turnover time of the supergranules (cf. Sect.~\ref{multi_scale_fluct}), and another at $\sim 35$~days that corresponds to an increase in the power level of the light fluctuations and a possible short-term cycle in the total area of the starspots (see below), respectively.

(iii) The multifractality of the Kepler-30 time series is principally due to the long-range correlations with a minor contribution from a broad non-Gaussian probability density distribution of the fluctuations. This result is found by comparing the original time series with their shuffled and phase-randomized surrogates. Our $h(t, \tau)$ plots suggest also that the small scale fluctuations that dominate the function $F_{q}$ for $q <0$ have a remarkable persistence, while the large scale fluctuations dominating for $q>0$ have a random and uncorrelated behaviour similar to that of a white noise, once the effect of the rotational modulation has been removed;

(iv) A maximum entropy spot modelling shows that the photospheric features of Kepler-30 evolve on timescales ranging from $10-20$ days for individual active regions up to a few hundreds of days for the longer-lived active longitudes. Their migration  can be used to estimate a lower limit for the relative surface shear of $\Delta \Omega / \Omega \sim 0.02 \pm 0.01$ that should be taken with great caution owing to the rapid evolution of the individual starspots that can mimic the effects attributed to differential rotation. A short-term cycle of about $\sim 34$ days in the total area of the starspots could be present and its timescale compares well with that found in the difference RTS timeseries as well as with a crossover (a slope change) in the fluctuation functions of the residuals of the spot modelling.

(v) Finally, we note a coincidental proximity of some activity timescales with the synodic periods of the two closest planets. The short-term spot cycle of  $\sim$34 days is close to the synodic period of 35.2 days of the planet Kepler-30b when a mean rotation period of $16$~days is adopted for Kepler-30. From the Lomb-Scargle periodogram of the RTS time series, a 23.1-day period emerges, that is close to the synodic period of Kepler-30c of 22 days. This period also appears in the multifractal analysis as a crossover of the fluctuation functions and is likely associated with the characteristic evolutionary timescales of active regions in Kepler-30 as indicated by the spot modelling. For the most distant planet Kepler-30d, we did not identify any similar coincidence. Nevertheless, the large separations of planets b and c suggest a great caution in the attribution of such coincidences to a possible star-planet magnetic interaction (see Sect.~\ref{spotted_area_var} for more detail).

We pointed out the relevant timescales and the persistence characteristics of the light fluctuations of Kepler-30, an active Sun-like star. This information can provide constraints for the models of the stellar flux variations based on the effects of magnetic fields and surface convection, thus promising to contribute to the refinement of those models in future investigations. 

As a perspective, our multifractal methods  are particularly interesting for analysing ongoing TESS and future PLATO high-precision photometric timeseries. They represent a useful complement to regularized spot models that can be used for an investigation of the activity phenomena on timescales comparable with the stellar rotation period or longer, while multi-fractal methods can cover the full range of timescales and characterize the persistence of the fluctuations produced by magnetoconvection on timescales remarkably shorter than the rotation period which are not accessible to spot modelling. On the other hand, we have illustrated how the two methodologies provide comparable results in the range of timescales they have in common. In this case, spot modelling allows us to gain more physical insight on the origin of the characteristic timescales pointed out by the multifractal analysis, at least when they can be associated with starspot evolution or activity cycles.  To reduce the degeneracies of spot modelling, it is advisable to target stars whose spin axis inclination can be constrained through the occultations of spots during planetary transits such as in the case of Kepler-30 \citep{sanchis} or by means of asteroseismology \citep[e.g.,][]{Ballotetal06}.

\begin{acknowledgements}
The authors are grateful to an anonymous referee for a careful reading of the manuscript and several valuable suggestions that helped them to improve their work. 
DBdeF acknowledges financial support from the Brazilian agency CNPq-PQ2 (grant No. 311578/2018-7). Research activities of STELLAR TEAM of Federal University of Cear\'a are supported by continuous grants from the Brazilian agency CNPq. AFL gratefully acknowledges support from the INAF Mainstream Project entitled ``Stellar evolution and asteroseismology in the context of PLATO space mission'' coordinated by Dr. Santi Cassisi. This paper includes data collected by the \textit{Kepler} mission. Funding for the \textit{Kepler} mission is provided by the NASA Science Mission directorate. All data presented in this paper were obtained from the Mikulski Archive for Space Telescopes (MAST). The authors would like to dedicate this paper to all the victims of the COVID-19 pandemic around the world.

\end{acknowledgements}


\begin{appendix}
\section{The multifractal background}
\label{multifractal_background}
In the present appendix, we present the steps of the MFDMA algorithm according to \citet{gu2010}:

\textbf{Step 1}: Construct the sequence of cumulative sums of time series $x(t)$ over time $t=1,2,3,...,N$, assuming the datapoints are evenly spaced:
\begin{equation}
\label{eq1}
y(t)=\sum^{t}_{i=1} x(i), \quad t=1,2,3,...,N,
\end{equation}
where $N$ is the total number of datapoints in the time series and the index $i$ is the time index $t$. 

\textbf{Step 2}: Calculate the moving average function of Eq. (\ref{eq1}) in a moving window:
\begin{equation}
\label{eq1a}
\tilde{y}(t)=\frac{1}{n}\sum^{\left\lceil n-1\right\rceil}_{k=0}y(t-k),
\end{equation}
where $n$ is the window size, and $\left\lceil (x)\right\rceil$ is the smallest integer that is not smaller than argument $(x)$\footnote{It is possible to adopt different kinds of moving averages instead of that defined by eq.~(\ref{eq1a}). For example, instead of considering datapoints that precede the given time $t$, it is possible to consider points that follow it or that are taken half before and half after the time $t$ \citep[see][]{gu2010}.};

\textbf{Step 3}: Detrend the series by removing the moving average function, $\tilde{y}(i)$, and obtain the residual sequence, $\epsilon(i)$, through:
\begin{equation}
\label{eq2}
\epsilon(i)=y(i)-\tilde{y}(i),
\end{equation}
where $ n \leq i \leq N$. 
We divide the residual series $\epsilon(i)$ into $N_{n}$=int$[N/n-1]$ non-overlapping segments of the same size $n$, where int$[x]$ is the largest integer that is not larger than $x$. Each segment can be denoted by $\epsilon_{\nu}$ so that $\epsilon_{\nu}(i)=\epsilon(l+i)$ where $1\le i \le n$ and $l=(\nu-1)n$;

\textbf{Step 4}: Calculate the root-mean-square (RMS) fluctuation function, $F_{\nu}(n)$, for a segment of size $n$:
\begin{equation}
\label{eq3}
F_{\nu}(n)=\left\{\frac{1}{n}\sum^{n}_{i=1}\epsilon^{2}_{\nu}(i)\right\}^{\frac{1}{2}}.
\end{equation}

\textbf{Step 5}: Generate the fluctuation function $F_{q}(n)$ of the $q$th order:
\begin{equation}
	\label{eq4}
	F_{q}(n)=\left\{\frac{1}{N_{n}}\sum^{N_{n}}_{\nu=1}F^{q}_{\nu}(n)\right\}^{\frac{1}{q}}, 
	\end{equation}
for all $q\neq 0$, where the $q$th-order function is the statistical moment (e.g., for $q$=2, we have the variance), while for $q=0$,
\begin{equation}
	\label{eq4b}
	\ln\left[F_{0}(n)\right]=\frac{1}{N_{n}}\sum^{N_{n}}_{\nu=1}\ln [F_{\nu}(n)].  
	\end{equation}
The scaling behaviour of $F_{q}(n)$ follows the relationship
\begin{equation}
	\label{eqh2}
	F_{q}(n)\sim n^{h(q)}, 
	\end{equation}
where $h(q)$ denotes the Holder exponent or generalized Hurst exponent. Each value of $q$ yields a slope $h$ and, in particular, $q=2$ gives the classical Hurst exponent.

\textbf{Step 6}: Knowing $h(q)$, the multifractal scaling exponent, $\tau(q)$, can be computed:
\begin{equation}
\label{eq5}
\tau(q)=q h(q)-1.
\end{equation}	

\textbf{Finally}, the singularity strength function, $\alpha(q)$, and the multifractal spectrum, $f(\alpha)$, are obtained via a Legendre transform:
\begin{equation}
\label{eq7}
\alpha(q)=\frac{d\tau(q)}{dq}
\end{equation}	
and
\begin{equation}
\label{eq6}
f(\alpha)=q\alpha-\tau(q).
\end{equation}	
By definition, for a monofractal signal, $h$ is the same for all values of $q$ and is equal to $\alpha$ giving a single-value spectrum $f(\alpha)=f(h)=1$, while for a multifractal signal, $h(q)$ is a function of $q$, and the multifractal spectrum is generally parabolic \citep[see Fig.~2 from][]{defreitas2017}. 

We use the following model parameters to compute the multifractal spectrum, as recommended by \cite{gu2010}: $N=30$; $q\in[-5,5]$ with a step size of 0.2; the lower bound of segment size $n$, which is denoted as $n_{\rm min}$ and set to 10; and the upper bound of segment size $n$, which is denoted as $n_{\rm max}$ and is given by \textbf{$N/10$}. An estimate of the standard deviation of the Hurst exponent is provided by \citet{Kantelhardt15} and is $<0.03$ when we consider timeseries with more than $10^{4}$ datapoints as in our case. 

\subsection{Multifractal indicators}\label{mi}
We use a set of four multifractal indicators that can be extracted from the quantities defined in Eqs. \ref{eq5}, \ref{eq7} and \ref{eq6}, in particular from the multifractal spectrum $f(\alpha)$ as illustrated in Fig. 2 of \cite{defreitas2017}. In this paper, we refer to this same figure to describe the shape of the multifractal spectrum. Here, we show a list of the main  indicators: 

\textbf{Parameter $\alpha_{0}$:}

The $\alpha_{0}$ parameter is the value of $\alpha$ corresponding the maximum of the multifractal spectrum $f(\alpha)$. It delivers information about the structure of the process producing the fluctuations, with a high value indicating that it is less correlated and has a fine structure \citep{Krzyszczak}. In addition, this parameter is strongly affected by signal variability. This is evidenced when one investigates the different sources of multifractality that are present in an astrophysical signal (see Sect.~\ref{origins_of_mf}). 

\textbf{Singularity parameter $\Delta f_{\rm min}(\alpha)$:} 

The parameter $\Delta f_{\rm min}(\alpha)$ characterizes the broadness, which is defined as the difference $f(\alpha_{\rm max})-f(\alpha_{\rm min})$ of the singularity spectrum. This difference provides an estimate of the spread in changes in fractal patterns. 
Consequently, if $\Delta f_{\rm min}(\alpha)$ is positive, the left-hand side of the spectrum is less deep, while a negative value indicates that this side is deeper. On the other hand, when $\Delta f_{\rm min}(\alpha)$ is null, the depths of the tails are the same on both sides. As quoted by \cite{ihlen}, larger fluctuations ($q>0$) imply that the singularities are stronger, whereas the smaller fluctuations ($q<0$) indicate that the singularities are weaker \citep[cf.][]{tanna}. In other words, multifractal spectra that have a longer right tail than the left one reveal that the structure of the time series is more regular and less dominated by extreme (maximal) values and, therefore, the parameter $\Delta f_{\rm min}(\alpha)$ can be a useful way to estimate the impact of noise in the periodic signal.

\textbf{Degree of asymmetry ($A$):}

This index is defined as the ratio:
\begin{equation}
\label{eq8}
A=\frac{\alpha_{\rm max}-\alpha_{0}}{\alpha_{0}-\alpha_{\rm min}},
\end{equation}
where $\alpha_{0}$ is the value of $\alpha$ where $f(\alpha)$ is maximal, while $\alpha_{\rm min}$ and $\alpha_{\rm max}$ are the minimum and maximum values of the singularity exponent $\alpha$ as defined by eq.~\ref{eq7}, respectively. The value of  $A$ indicates one of three possible skewness of the singularity spectrum: right-skewed ($A>1$), left-skewed ($0<A<1$) or symmetric ($A=1$). 

\textbf{Degree of multifractality ($\Delta \alpha$):}

This index represents the measure of the interval:
\begin{equation}
\label{eq9}
\Delta \alpha=\alpha_{\rm max}-\alpha_{\rm min},
\end{equation}
where $\alpha_{\rm max}$ and $\alpha_{\rm min}$ are defined above. A low value of $\Delta\alpha$ indicates that the time series is close to fractal with the multifractal strength being  higher when $\Delta\alpha$ increases \citep{defreitas2009,defreitas2017}. As mentioned by \cite{defreitas2016}, larger values of $\Delta\alpha$ denote more complex fluctuations, whereas smaller values indicate that the spectrum tends towards the monofractal limit. According to \cite{makowiec}, if $\Delta\alpha$ is less than 0.05 a monofractal behaviour of the spectrum should be assumed because of the intrinsic precision in deriving such a parameter from a statistics based on a number of datapoints that is in any case limited.  

\begin{table*}
\caption{The multifractal indicators extracted from the standard MFDMA method for PDC, SAP and RTS. The results are summarized considering the Original (O), Shuffled (S) and Phase randomized (P) data. From left to right, we have: the columns $q<0$ and $q>0$ (see Section 4.2 for further details), the multifractality due to long-term correlation is stronger than that due to non-linearity; singularity parameter; parameter $\alpha_{0}$; global Hurst exponent; degree of multifractality and; degree of asymmetry.}              
\label{tab1}      
\centering                                      
\begin{tabular}{c c c c c c c c c c c c c c c c c c c c}          
\hline\                        
	&	$q<0$	&	$q>0$	&	$\Delta f_{min}^{O}(\alpha)$	&	$\alpha_{0}^{O}$	&	$\alpha_{0}^{S}$	&	$\alpha_{0}^{P}$	&	$H_{O}$	&	$H_{S}$	&	$H_{P}$	&	$\Delta \alpha_{O}$	&	$\Delta \alpha_{S}$	&	$\Delta \alpha_{P}$	&	$A_{O}$	&	$A_{S}$	&	$A_{P}$	\\
\hline
PDC	&	1	&	1	&	0.59	&	0.59	&	0.56	&	0.60	&	0.42	&	0.53	&	0.44	&	0.97	&	0.12	&	0.86	&	2.95	&	1.13	&	3.20	\\
SAP	&	1	&	1	&	0.67	&	0.64	&	0.52	&	0.66	&	0.47	&	0.55	&	0.47	&	0.90	&	0.06	&	0.92	&	2.88	&	1.79	&	2.64	\\
RTS	&	1	&	1	&	0.52	&	0.81	&	0.51	&	0.75	&	0.58	&	0.49	&	0.61	&	1.12	&	0.13	&	0.73	&	2.47	&	1.83	&	3.50	\\
\hline                                             
\end{tabular}
\end{table*}

\subsection{Results based on the standard multifractal approach}
\begin{figure}
\begin{center}
	\includegraphics[width=0.90\columnwidth]{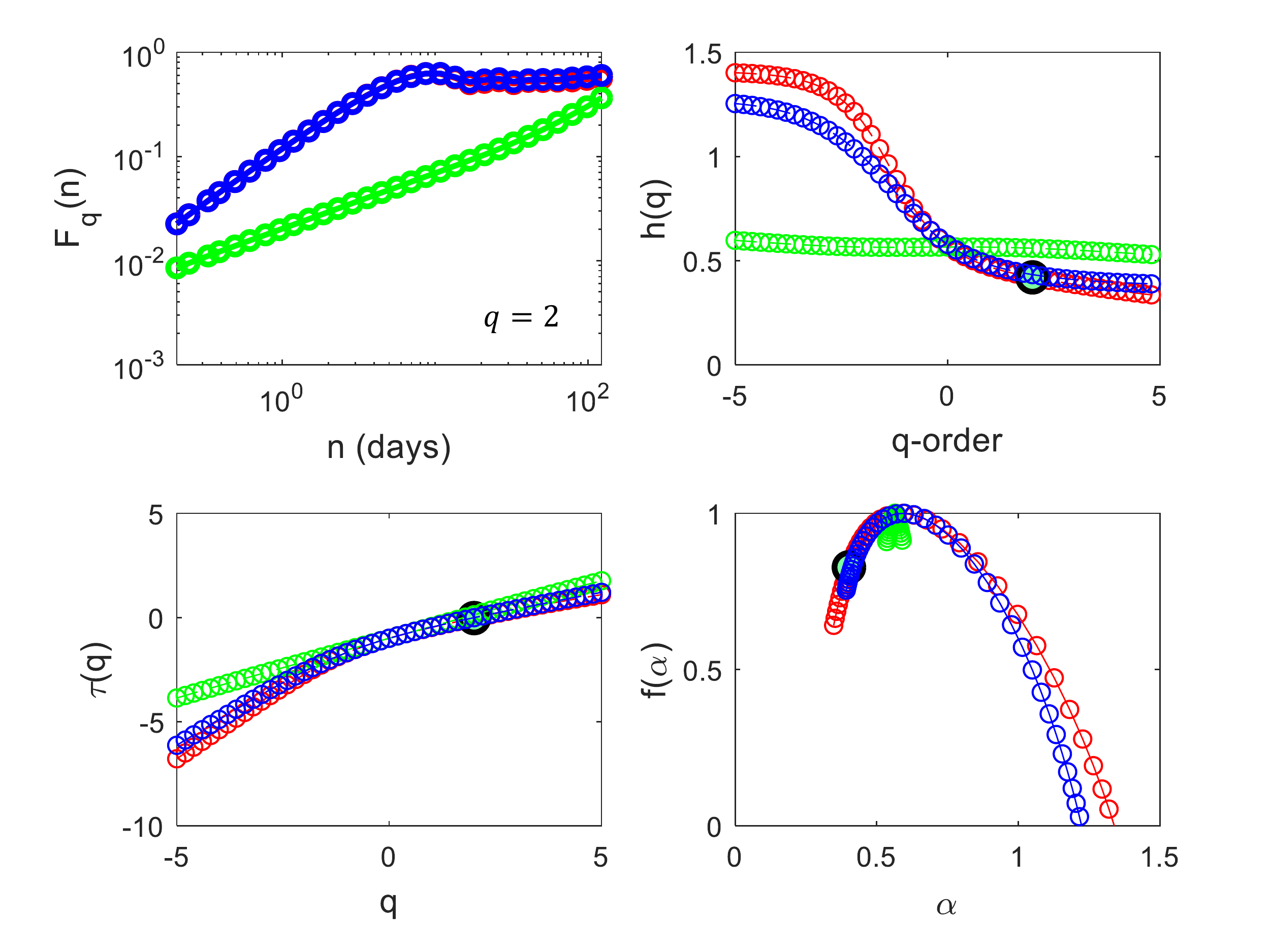}
	\end{center}
	\caption{Multifractal analysis for PDC time series following steps 5 and 6 presented in Section 4. \textit{Top panel}: The original (in red), the shuffled (green) and phase randomized (blue) data are based on the procedure mentioned in Section 4. \textit{Left top}: the multifractal fluctuation function $F_{q}(n)$ obtained from MFDMA method for only $q=2$, indicated henceforth as a big circle. \textit{Right top}: $q$-order Hurst exponent ($h(q)$) as a function of $q$-parameter. This panel shows the truncation originated from the leveling of the $h(q)$ for positive $q$'s. \textit{Left bottom}: comparison of the multifractal scaling exponent $\tau(q)$ of three data. In this panel is possible to identify a crossover in $q\sim -1$. \textit{Right bottom}: multifractal spectrum $f(\alpha)$ of three time series, respectively.}
	\label{figMFDMA1}
\end{figure}

\begin{figure}
\begin{center}
\includegraphics[width=0.45\textwidth]{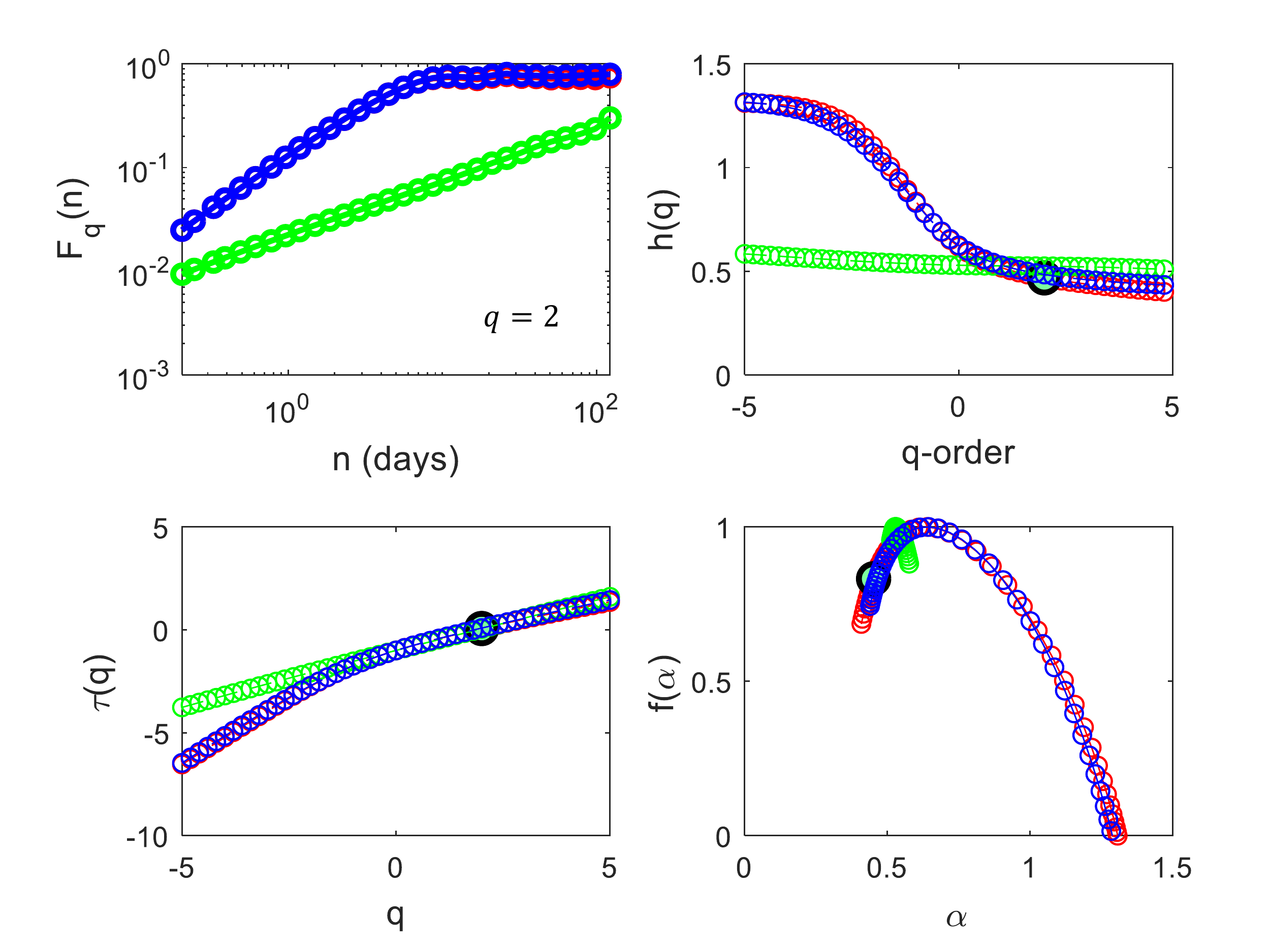}
\end{center}
\caption{Idem figure \ref{figMFDMA1} for SAP time series.}
\label{figMFDMA2}
\end{figure}

\begin{figure}
\begin{center}
\includegraphics[width=0.45\textwidth]{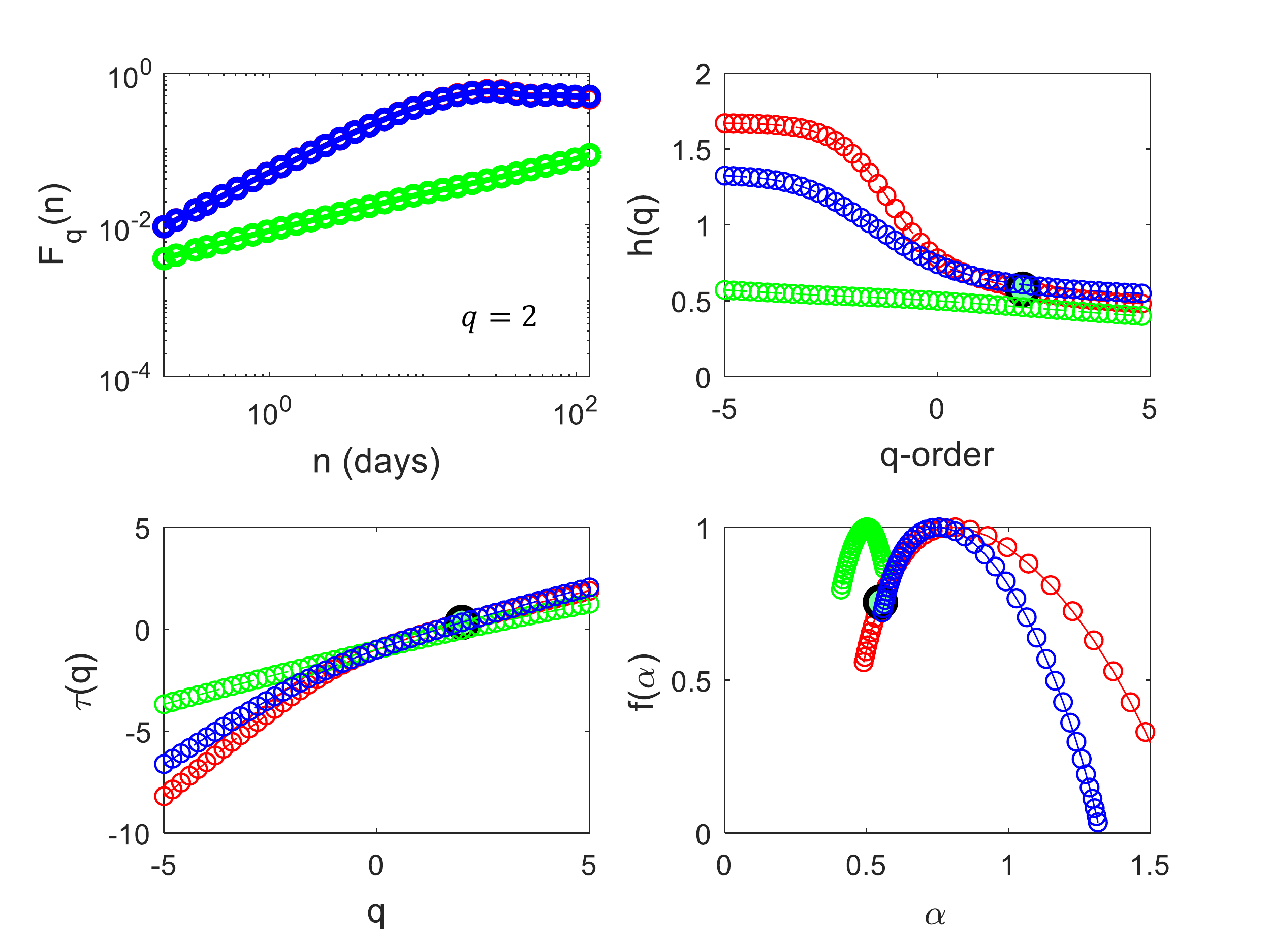}
\end{center}
\caption{Idem figure \ref{figMFDMA1} for RTS data.}
\label{figMFDMA3}
\end{figure}

First, we calculate the MFDMA fluctuation functions $F_{q}(n)$ as a function of window size $n$ (in days) for the three time series PDC, SAP, RTS. The scale parameter $n$ is varied from 10 to $N/10$, and the exponent $q$ is varied from $-$5 to 5 in steps of 0.2.  The scaling pattern of $F_{q}(n)$ of original data (red lines) for $q = 2$ is shown in the top left panels  of Figs. \ref{figMFDMA1}, \ref{figMFDMA2} and \ref{figMFDMA3}. We  repeat the analysis for a set of 200 randomly shuffled series as well as for 200 phase-randomized series (blue and green lines, respectively).

\begin{figure}
\begin{center}
\includegraphics[width=0.45\textwidth]{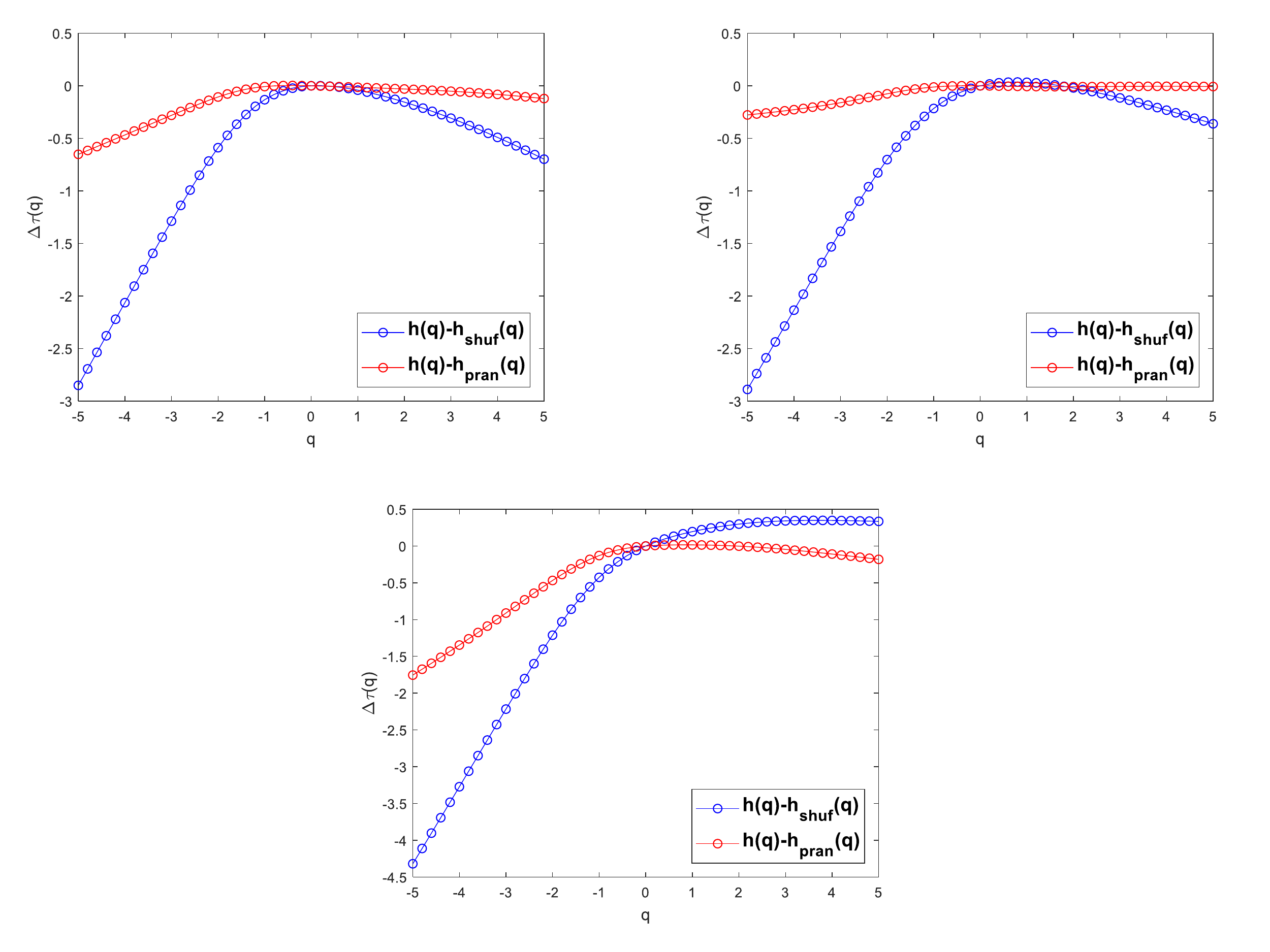}
\end{center}
\caption{Deviations of scaling exponent for (top left) PDC, (top right) SAP and (bottom) RTS data.}
\label{figDtau}
\end{figure}

For PDC, SAP and RTS from Kepler-30, the standard multifractal characterization was performed following the procedures described above and also for the series obtained from surrogates of the original data sets. For all of the timeseries, the above four multifractal indicators together with the Hurst exponent $H$ were obtained, and their values are shown in Table 1 where the parameters associated with the original timeseries are denoted with an (O) superscript, while those associated with the shuffled and the phase-randomized series with an (S) or (P) superscripts, respectively.

For all the series and $q$-domains, the multifractality due to correlation is stronger than that due to the non-Gaussianity of the distribution of the fluctuations with fat tails as indicated by ``1'' and the positive values of $\Delta f_{\rm min}^{O}(\alpha)$. This value does not imply that there are only long-term correlations in fluctuations, but it implies that non-linearities due to fat-tailed distributions are very weak. As a consequence, a time-independent structure with $h(q)\approx 0.5$ is shown in the shuffled time series (see Figs. \ref{figMFDMA1}, \ref{figMFDMA2} and \ref{figMFDMA3}). After shuffling, all time series exhibit a smaller degree of multifractality than the original one. This result can be emphasized by the multifractal spectra (green curves from right top panels). In general, this analysis is not conclusive for explaining the source of this behaviour. In this case, it is necessary to verify the behaviour of $\tau(q)$. As illustrated by Fig. \ref{figDtau}, there is a strong dependence of $h(q)-h_{shuf}(q)$ on $q$, which is clearest for $q<0$, where a deeper right tail occurs in multifractal spectra (see right bottom panels from Figs. \ref{figMFDMA1}, \ref{figMFDMA2} and \ref{figMFDMA3}). In contrast, the dependence on $q$ can be neglected for the phase-randomized data, except for $q<0$ in PDC and RTS data as illustrated by Fig. \ref{figDtau}. Thereby, the two sources of multifractalities appear in $q<0$, the domain where the rotational modulation occurs. Basically, the correlations and non-linearities are negligible for $q>0$, where the strong fluctuations due to the noise (correlated or not) appear.

Because the presence of rotational modulation and noise can affect the multifractal indicators, we decided to investigate the changes of the multifractal spectrum and compare the indicators calculated from the original series with those obtained from the surrogate series. Consequently, it is possible to find the source(s) that affect the values of $\alpha_{0}$, $H$, $\Delta\alpha$ and $A$ following the behaviour of their calculated values from the surrogate series.

Firstly, the relation between the values of $\alpha_{0}$ in the three time series show a clear difference among them. It is interesting to note that SAP data have larger $\alpha^{O}_{0}$ values than the PDC time series, which means that in PDC data the fluctuations governing the rotational modulation are more correlated and have a less fine structure (a structure more regular in appearance) compared to the fluctuations of the SAP data. For RTS data, the fluctuations are even less correlated and have a more fine structure (a structure less regular in appearance) compared to the processes governing the PDC and SAP data. Secondly, the values of $H$ for the surrogate data indicate that shuffling the data destroys the correlations, and therefore $H$ tends to 0.5, whereas the phase-randomized data recover a value very close to that found for the original series.

In contrast, the degree of multifractality $\Delta\alpha$ changes more broadly. Because this parameter is connected to the richness of the data structure, we highlight that the $\Delta\alpha$ for the original time series of RTS indicates such data may promote some values of the fluctuations, making the signal structure richer. Nevertheless, there is an important detail. As $\Delta f_{\rm min}^{O}(\alpha)$ is the smallest value, the broadness of  $\Delta\alpha$ is a mixing between strong and weak fluctuations, and therefore the noise has an important effect over the data, as can be emphasized by the small values of $\Delta\alpha_{S}$. However, following the criterion of \cite{makowiec}, the shuffling process does not reduce the series to a monofractal. 

It is possible to observe a dissimilarity for the asymmetry parameter $A_{O}$ among the time series for both the original and the surrogate data. As shown in Figs. \ref{figMFDMA1}, \ref{figMFDMA2} and \ref{figMFDMA3}, and Table~\ref{tab1}, it is interesting that generally the spectra are rather right-skewed (which suggests that fine structures are more frequent) or tend to be symmetrical in shape. However, the extreme events become stronger for the RTS data, as the left side of the spectrum of this series is deeper than those from PDC and SAP data.

In conclusion, it is interesting to address a comparison of the parameters of the multifractal spectra between the two time series, namely PDC and SAP. It allows us to address the question of whether the MFDMA method can be applied as an indicator of the changes in the dynamics of fluctuations and, therefore, of processes occurring in the stellar atmosphere from which the stellar flux comes. It can be observed that the values of $\alpha_{0}$ change only slightly and are greater in the SAP data. On the other hand, the degree of multifractality is more developed in the PDC data, whereas the asymmetry for SAP time series is slightly more positive. Some differences between both time series can be also seen when analysing the absolute differences of Hurst exponents for the original and shuffled data or original and surrogates (see Fig. \ref{figDtau}). Even though we do not see a change between both time series, it is noticed that the source of multifractality due to the contribution of long-range correlations is dominant.

\section{Fluctuation functions for a sinusoidal signal}
\label{fluc_func_sinusoid}

\begin{figure}
\begin{center}
\includegraphics[width=9cm, angle =0]{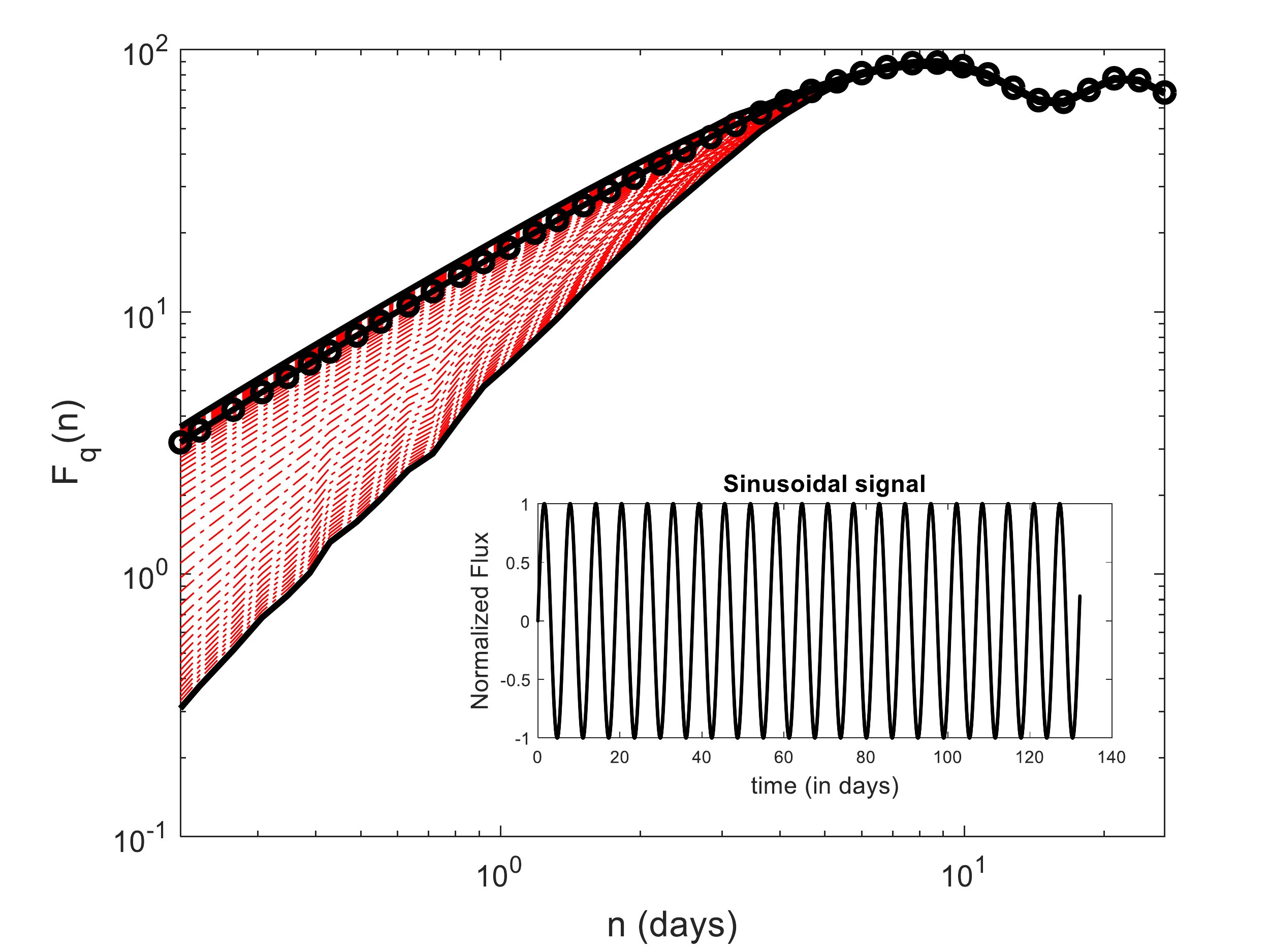}
\end{center}
\caption{The log-log plot of fluctuation functions $F_{q}(n)$ vs. $n$ for a sinusoidal time series (shown in the subplot) with period of $\sim$ 6.3 days.}
\label{figSine}
\end{figure}

We consider a noiseless purely sinusoidal signal with constant amplitude and phase to study how the fluctuation functions computed with the MFDMA behave under the effect of a sinusoidal modulation. \citet{sps2009} investigated a similar case using a re-scaled range analysis, the so-called $R/S$ method \citep[e.g.,][]{Kantelhardt15}, to evaluate the fluctuations function, but did not discuss the case of the MFDMA method. In our case, we see that the $F_{q}(n)$ fluctuation functions, plotted as a function of $n$ (see Fig~\ref{figSine}) show a linear trend followed by an oscillating phase, a behaviour similar to that shown in Figures~\ref{figFlu1}, \ref{figFlu2} and \ref{figFluResidual}. Other factors can lead to deviations from this ideal  behaviour, such as noise and/or the superposition of other low-amplitude periodicities. The combination of these contributions should make the simulated time series closer to the observed ones. 

\section{Spot modelling}
\label{appendix_spot_model}
To map the photosphere of Kepler-30,  we subdivide it into 200 surface elements of side $18^{\circ} \times 18^{\circ}$ and assume that in each element there are dark spots with a filling factor $f$, solar-like faculae with a filling factor $Qf$, and unperturbed photosphere with a filling factor $1-(Q+1)f$, where $Q$ is the ratio of the area of the faculae to that of the spots in each element. For simplicity, $Q$ is assumed constant in our model.  

The contrast of the dark spots is defined as $c_{\rm s} \equiv I_{\rm spot}/I_{\rm u}$, where $I_{\rm spot}$ is the brightness of the spotted photosphere and  $I_{\rm u}$ that of the unperturbed photosphere; $c_{\rm s}$ is assumed to be constant. In our case, $c_{\rm s}=0.85$ in the Kepler passband has been derived from the occulted spots producing characteristic bumps along the photometric profiles of the transits \citep{sanchis}. 

The facular contrast is assumed to be zero at the centre of the stellar disc and maximum at the limb as we observe in the Sun, that is, $c_{\rm f} = c_{\rm f 0} (1- \mu)$, where $c_{\rm f0} = 1.115$ is the contrast at the limb and $\mu \equiv \cos \psi$, where $\psi$ is the angle between the normal to the surface at a given point and the line of sight.  Note that the effect of the faculae is parametrized by the product $Qc_{\rm f0}$, thus, their contrast and their area ratio are not independent parameters \citep[cf.][]{Lanza16}. Therefore, we keep $c_{\rm f 0}$ constant and very $Q$ (see below).

The limb-darkening of the unperturbed photosphere is expressed by means of a quadratic law:
\begin{equation}
I_{\rm u} (\mu) \propto a_{\rm p} + b_{\rm p} \mu + c_{\rm p} \mu^{2},
\end{equation}
where $a_{\rm p}$, $b_{\rm p}$, and $c_{\rm p}$ are derived from the limb-darkening coefficients given by \citet{sanchis}. 

Our model assumes that the distribution of the filling factor $f$ over the surface of the star is fixed while fitting the light curve. Because spots are evolving, this means that we cannot apply our model to fit the entire photometric time series, but we must fit individual intervals of duration $\Delta t_{\rm f}$ over which the hypothesis of a fixed spot pattern is satisfied. In stars with a slowly evolving spot pattern, the best choice is to take $\Delta t_{\rm f}$ equal to the mean rotation period. This will give a uniform sampling of all the longitudes along each individual rotation \citep{Lanzaetal07}. 

Unfortunately, this is not the case for Kepler-30 because its starspots evolve quite rapidly, thus imposing a shorter time interval to adequately fit its light modulation. Following the method applied in previous modelling of other CoRoT or Kepler targets \citep[e.g.][]{lanza2019}, we derive the best values of $\Delta t_{\rm f}$ and $Q$ by applying a simplified spot model assuming only three discrete spots plus a uniform background \citep{lanza}. 

The value of $\Delta t_{\rm f}$ is derived by considering that in general the chi square of the best fit to the entire light curve decreases  with the decrease of $\Delta t_{\rm f}$. However, the reduction of the chi square becomes small when $\Delta t_{\rm f}$ becomes equal to or shorter than the typical timescale of spot evolution. Therefore, we progressively decrease $\Delta t_{\rm f}$ until the decrease of the chi square is no longer significant according to a statistical test based on the Fisher-Snedecor statistics and in this way we select the optimal $\Delta t_{\rm f}$. Note that for each trial value of $\Delta t_{\rm f}$, we compute models with different values of $Q$ spanning a large interval to select the minimum of the chi square. This is made possible by the relatively short CPU time required to compute the best fit with the three-spot model in comparison to the full model assuming a continuous spot distribution. 

The value of $Q$ minimizing the chi square for the optimal value of $\Delta t_{\rm f}$ is adopted for our spot modelling. In the case of Kepler-30, we find that the optimal $\Delta t_{\rm f} = 11. 963$~days, that is, $\sim 0.75$ of the mean rotation period, while $Q=0.5$, that is much smaller than the value adopted for the spot modelling of the Sun as a star by \citet{Lanzaetal07}, where $Q_{\odot} = 9.0$. This is in agrement with the finding that dark spots dominate the optical light modulation of stars significantly more active than the Sun as is the case for Kepler-30 \citep[cf.][]{Radicketal18}. 

The inclination of the stellar spin to the line of sight is derived by assuming that the stellar spin is normal to the orbital plane of the planet Kepler-30c that is the largest body in the system after the star. Nevertheless, the orbits of the other planets are virtually coplanar, so this is equivalent to assuming that the stellar spin is perpendicular to the mean plane of the planetary orbits. Note that the possibility of constraining the inclination of the stellar spin reduces the degeneracies of our spot modelling and is made possible by the presence of transiting planets around the star. 

The parameters adopted in our spot modelling are listed in Table~\ref{table_spot_model}.  The mass and the radius of the star together with its rotation period are used to compute the surface gravity as a function of the latitude to account for the effects of the gravity darkening. They are of the order of $10^{-6}$ in relative flux units and do not affect our solution. 

The effects of the uncertainties in the model parameters on the spot modelling have been discussed in detail by \citet{Lanzaetal09} e \citet{lanza2019}. They are negligible for our application because we are mainly interested in the typical timescales of starspot evolution rather than on the absolute values of the spotted area or of the surface differential rotation.
\begin{table}
\begin{center}
\begin{tabular}{lrr}
\hline
& & \\ 
Parameter  & & Reference \\
& & \\ 
\hline
& & \\
Mass (M$_{\odot}$) & 0.99 & 1 \\
Radius (R$_{\odot}$)  & 0.95 & 1 \\
$P_{\rm rot}$ & 16.0 & 1\\
$i$ (deg) & 89.68 & 2 \\ 
$c_{\rm s}$ & 0.85 & 1 \\
$a_{\rm p}$ & 0.22 & 1\\
$b_{\rm p}$ & 1.18 & 1 \\
$c_{\rm p}$ & -0.40 & 1 \\
$Q$ & 0.5 & 2 \\
$\Delta t_{\rm f}$ (d)  & 11.963 & 2 \\
 & & \\
\hline
\end{tabular}
\caption{Parameters adopted in the spot modelling of Kepler-30. References: 1: \citet{sanchis}; 2: present work}
\label{table_spot_model}
\end{center}
\end{table}

\begin{figure}
\begin{center}
\includegraphics[width=0.35\textwidth,trim={2.5cm 3.5cm 2.5cm 2.5cm}]{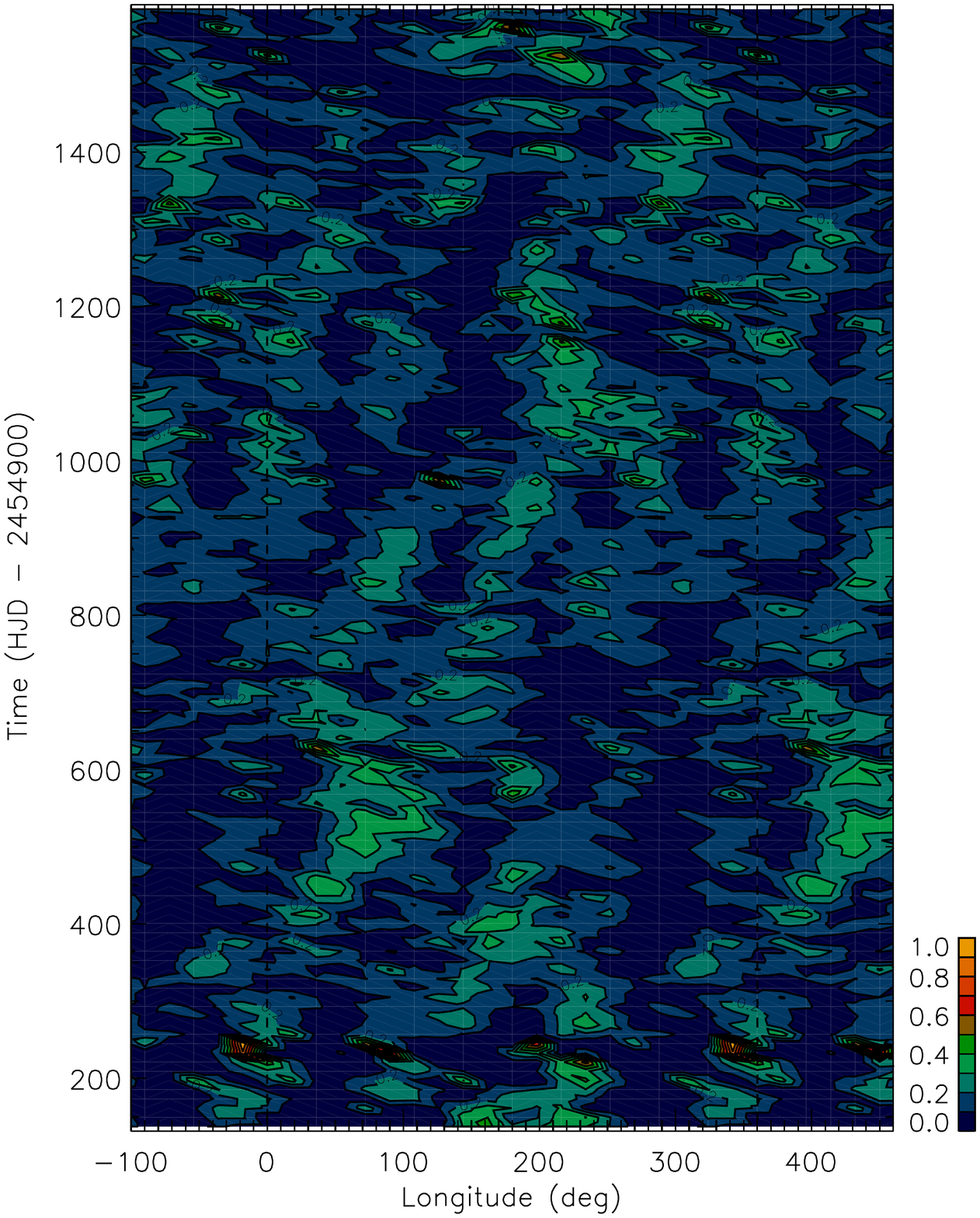}
\end{center}
\caption{Distribution of the spot filling factor vs. the longitude and time for the ME regularized spot model  of the SAP light curve of Kepler-30. The minimum of the filling factor corresponds to dark blue regions, while the maximum is rendered in orange (see the colour scale close to the lower right corner of the plot). Note that the longitude scale of the horizontal axis is extended beyond the $[0^{\circ}, 360^{\circ}]$ interval to better follow the migration of the starspots.}
\label{spot_map_sap}
\end{figure}
\begin{figure}
\begin{center}
\includegraphics[width=0.34\textwidth,trim={2.5cm 3.5cm 2.5cm 4.5cm}, angle =90]{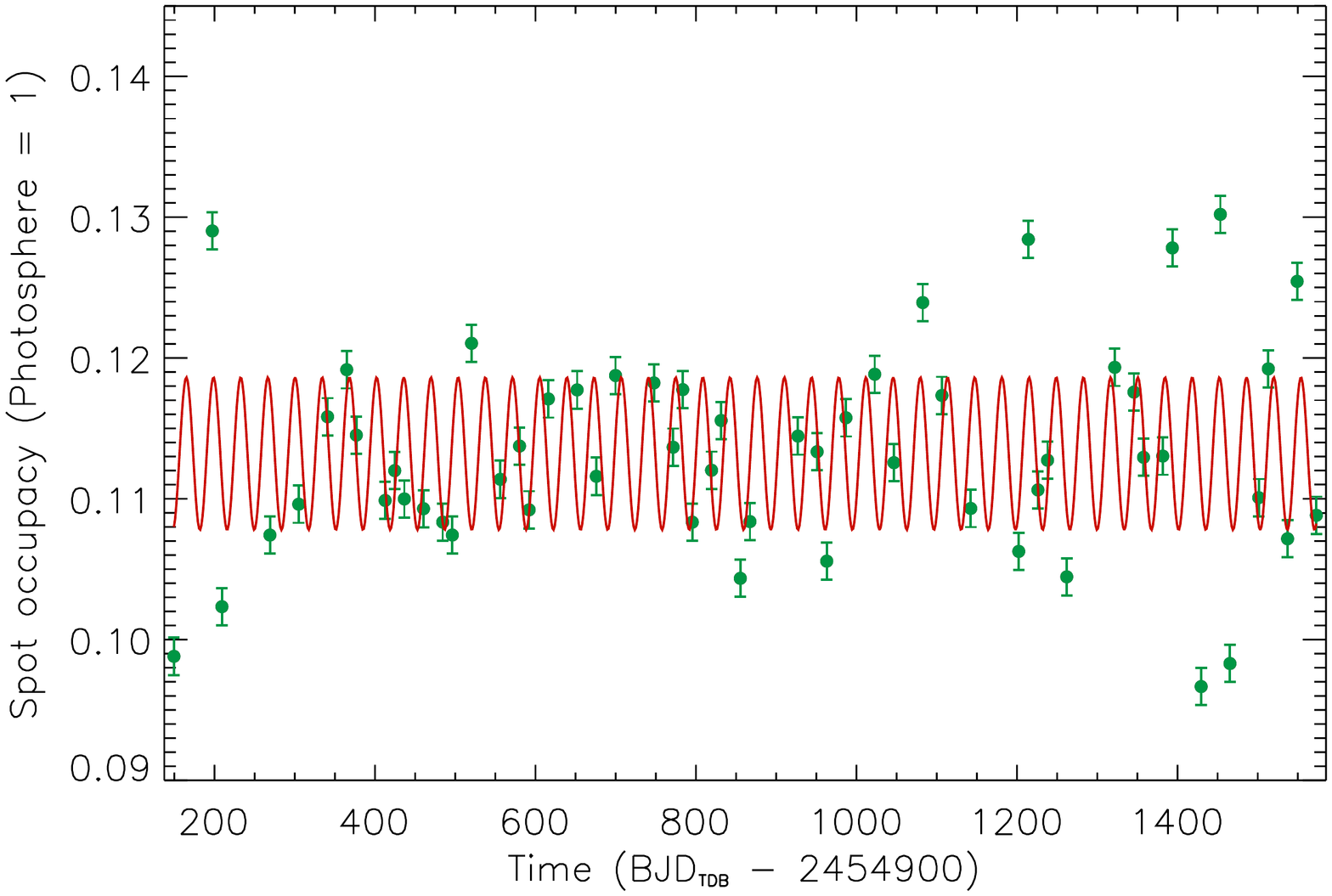}
\end{center}
\caption{The total spotted area on Kepler-30 as derived from the spot modelling of the SAP light curve vs. the time (green dots). A sinusoid with a period of 33.878~days corresponding to the maximum of the GLS periodogram is overplotted (red solid line).}
\label{total_spot_area_sap}
\end{figure}

By minimizing the chi square of the fit to the light curve with our continuous spot model, we can find a unique spot map for Kepler-30, but such a map is unstable in the sense that a small change in the data produces a large change in the map because most of the map details come from fitting the photometric noise and the evolution of small spots along each $\Delta t_{\rm f}$ time interval. Regularization can provide a unique map by reducing the fitting of the noise and of the small fluctuations and therefore the consequent instability. This is achieved by imposing a priori assumptions on the spot map that are coded into a regularization functional $S({\bf f})$ that depends on the vector ${\bf f}$, the components of which are the filling factors of the 200 surface elements of our map \citep[see eq.~(3.25) in][for the expression of $S$]{Lanza16}. 

In the case of a regularized best fit, instead of minimizing the chi square $\chi^{2}$ between the observed and the modelled fluxes, we minimize a linear combination of the $\chi^{2}$ and the regularizing functional $S$, viz:
\begin{equation}
Z({\bf f}) = \chi^{2}({\bf f}) -\lambda_{\rm ME} S({\bf f}), 
\end{equation}
where $\lambda_{\rm ME} > 0$ is a Lagrangian multiplier that rules the level of regularization. When $\lambda_{\rm ME}$ is zero, we get the unregularized model that has the minimum $\chi^{2}$ and residuals that are symmetrically  distributed around a zero mean $\mu_{\rm res}$. When we increase $\lambda_{\rm ME}$, we increase the $\chi^{2}$ above its minimum and the mean of the residuals is no longer zero because the ME regularization drives the map towards an unspotted photosphere, thus making the residuals systematically negative. In other words, the functional $S$ is designed to be maximal when the star is unspotted, so that, by introducing the regularization, we drive the map towards that of an unspotted star. 

The criterion to fix the Lagrangian multiplier is crucial to decide when the process of regularization has to be stopped to avoid that the $\chi^{2}$ becomes too large and the fit unacceptable. As in the case of similar stars with a photon shot noise of the single datapoint of the order of $\sim 1$\% of the amplitude of the flux modulation produced by starspots, we iteratively increase $\lambda_{\rm ME}$ until the absolute value of the mean of the residuals $| \mu_{\rm res}|$ becomes equal to the standard error $\epsilon_{\rm st}$ of the datapoints in the fitted interval of duration $\Delta t_{\rm f}$. The standard error is computed from the standard deviation $\sigma_{0}$ of the residuals of the unregularized best fit and the number of datapoints $M$ in each interval $\Delta t_{\rm f}$ as $\epsilon_{\rm st} = \sigma_{0}/\sqrt{M}$. For each interval $\Delta t_{\rm f}$, we determine $\lambda_{\rm ME}$ by enforcing the equality $ |\mu_{\rm res}| = \epsilon_{\rm st}$ within 5\% both for the fit of the PDC and the SAP light curves. 

The unregularized best fit of the SAP light curve obtained with our model is very similar to that of the PDC timeseries and is not shown here. The regularized spot map is shown in Fig.~\ref{spot_map_sap} and is similar to that obtained with the PDC light curve. The SAP light curve shows a range of amplitudes of the light modulation  larger than  the PDC light curve because the Kepler pipeline tends to reduce the variability at timescales longer than $15-20$ days.   As a consequence, the SAP spot map shows a wider distribution of the filling factor than the PDC map, but the location of the spots and their evolution are very similar to those shown in the PDC map. 

The  total spotted area as derived by the ME regularized spot model of the SAP light curve is plotted in Fig.~\ref{total_spot_area_sap} as a function of the time. As in the case of the PDC light curve, only the intervals $\Delta t_{\rm f}$ with datapoints having a sufficiently uniform distribution are considered to avoid systematic effects produced by the ME regularization (see Sect.~\ref{spotted_area_var}).  A modulation with a period of $\sim 33.9$ days is apparent as indicated by the sinusoid superposed to the plot.

\end{appendix}

\end{document}